\documentclass[twocolumn,showpacs,amsmath,amssymb,superscriptaddress]{revtex4}
\usepackage{graphicx}

\newlength{\onecolfig}
\newlength{\twocolfig}
\newcommand{\ion}[2]{\mbox{$^{#2}$#1$^+$}}
\newcommand{\CaI}[1]{\mbox{$^{#1}$Ca}}
\newcommand{\CaII}[1]{\ion{Ca}{#1}}
\newcommand{\lev}[3]{\mbox{$^{#1}$#2$_{\mbox{\tiny$#3$}}$}}
\newcommand{\hfslev}[3]{\mbox{#1$^{\mbox{\tiny$#3$}}_{\mbox{\tiny$#2$}}$}}

\newcommand{\unit}[1]{\,\mbox{#1}}
\newcommand{\kHz}{\unit{kHz}}
\newcommand{\MHz}{\unit{MHz}}
\newcommand{\GHz}{\unit{GHz}}

\newcommand{\torr}{\unit{torr}}
\newcommand{\mW}{\unit{mW}}
\newcommand{\uW}{\unit{$\mu$W}}
\newcommand{\mrad}{\unit{mrad}}
\newcommand{\mm}{\unit{mm}}
\newcommand{\um}{\unit{$\mu$m}}
\newcommand{\nm}{\unit{nm}}
\newcommand{\K}{\unit{K}}

\newcommand{\us}{\unit{$\mu$s}}
\newcommand{\ns}{\unit{ns}}

\newcommand{\degree}{\mbox{$^{\circ}$}}
\newcommand{\degC}{\mbox{\degree{}C}}

\newcommand{\s}{\unit{s}}

\newcommand{\citesec}[2]{\cite[\S{}#2]{#1}}   % {} to eliminate gap?
\newcommand{\etal}{{\em et al.}}

\newcommand{\ie}{{\em i.e.}}

\newcommand{\ish}{\mbox{$\sim$}\,}
\newcommand{\ltish}{\protect\raisebox{-0.4ex}{$\,\stackrel{<}{\scriptstyle\sim}\,$}}
\newcommand{\gtish}{\protect\raisebox{-0.4ex}{$\,\stackrel{>}{\scriptstyle\sim}\,$}}

\newcommand{\wee}[2]{\mbox{$\frac{#1}{#2}$}}
\newcommand{\sub}[1]{\mbox{$_{\mbox{\tiny #1}}$}}
\newcommand{\diff}[1]{\mbox{\/d$#1$}}

\newenvironment{centre}{\begin{center}}{\end{center}}

\begin{document}
\bibliographystyle{apsrev}

\title{Isotope-selective photo-ionization for calcium ion trapping}

\author{D. M. Lucas}
\affiliation{Department of Physics, University of Oxford, Clarendon Laboratory, Parks Road, Oxford OX1 3PU, U.K.}
\author{A. Ramos}
\affiliation{Department of Physics, University of Oxford, Clarendon Laboratory, Parks Road, Oxford OX1 3PU, U.K.}
\author{J. P. Home}
\affiliation{Department of Physics, University of Oxford, Clarendon Laboratory, Parks Road, Oxford OX1 3PU, U.K.}
\author{M. J. McDonnell}
\affiliation{Department of Physics, University of Oxford, Clarendon Laboratory, Parks Road, Oxford OX1 3PU, U.K.}
\author{S. Nakayama}
\affiliation{Department of Information and Computer Science, Kagoshima University, 1-21-40 Korimoto, Kagoshima 890-0065, JAPAN}
\author{J.-P. Stacey}
\affiliation{Department of Physics, University of Oxford, Clarendon Laboratory, Parks Road, Oxford OX1 3PU, U.K.}
\author{S. C. Webster}
\affiliation{Department of Physics, University of Oxford, Clarendon Laboratory, Parks Road, Oxford OX1 3PU, U.K.}
\author{D. N. Stacey}
\affiliation{Department of Physics, University of Oxford, Clarendon Laboratory, Parks Road, Oxford OX1 3PU, U.K.}
\author{A. M. Steane}
\affiliation{Department of Physics, University of Oxford, Clarendon Laboratory, Parks Road, Oxford OX1 3PU, U.K.}

\date{\today}

\begin{abstract}
We present studies of resonance-enhanced photo-ionization for isotope-selective loading of \CaII{} into a Paul trap.
The 4s$^2\,\lev{1}{S}{0}\leftrightarrow$ 4s4p\,\lev{1}{P}{1} transition of neutral calcium is driven by a 423\nm\ laser
and the atoms are photo-ionized by a second laser at 389\nm. Isotope-selectivity is achieved by using crossed atomic
and laser beams to reduce the Doppler width significantly below the isotope shifts in the 423\nm\ transition. The
loading rate of ions into the trap is studied under a range of experimental parameters for the abundant isotope
\CaII{40}. Using the fluorescence of the atomic beam at 423\nm\ as a measure of the Ca number density, we estimate a
lower limit for the absolute photo-ionization cross-section. We achieve loading and laser-cooling of all the naturally
occurring isotopes, without the need for enriched sources. Laser-heating/cooling is observed to enhance the
isotope-selectivity. In the case of the rare species \CaII{43} and \CaII{46}, which have not previously been
laser-cooled, the loading is not fully isotope-selective but we show that pure crystals of \CaII{43} may nevertheless
be obtained. We find that for loading \CaII{40} the 389\nm\ laser may be replaced by an incoherent source.
\end{abstract}


\maketitle

\section{Introduction}

Resonance-enhanced photo-ionization for loading ion traps was first demonstrated with magnesium and calcium ions by
Kj\ae{}rgaard \etal~\cite{00:Kjaergaard}. The same group has recently used crossed-beam Doppler-free excitation of the
272\nm\ 4s$^2\,\lev{1}{S}{0}\leftrightarrow$ 4s5p\,\lev{1}{P}{1} calcium transition, followed by photo-ionization, to
load selectively all the naturally occurring isotopes of this element; the loading was not observed directly, but by
using charge-exchange to infer the presence of the other isotopes from the fluorescence of laser-cooled \CaII{40} ions
which replaced them~\cite{02:Mortensen}. Gulde \etal\ showed that, using the 4s$^2\,\lev{1}{S}{0}\leftrightarrow$
4s4p\,\lev{1}{P}{1} Ca transition at 423\nm, followed by excitation close to the continuum by an ultraviolet photon,
photo-ionization is around five orders of magnitude more efficient than conventional electron bombardment
ionization~\cite{01:Gulde}.

As pointed out by these authors, photo-ionization has a number of advantages over electron bombardment. Only the
desired species is loaded into the ion trap, allowing pure crystals of particular isotopes to be obtained. Since no
electron beam is involved, there is no charging of insulating parts of the trap structure, which leads to drifting
electric fields that for many experiments must be accurately compensated. The efficiency of the photo-ionization
process allows much lower number densities of neutral atoms in the interaction region, greatly reducing the quantities
of material sputtered onto the trap electrodes; it has been shown that clean electrode surfaces reduce the heating rate
of trapped ions from the motional ground state~\cite{00:Turchette}, an important consideration for quantum logic
experiments in ion traps. For experiments in which high-finesse optical cavities are combined with ion traps, the lower
atomic beam density reduces the degradation of the mirror surfaces and can avoid the need for a separate ``loading
trap''~\cite{03:Mundt,03:Keller}.

Our particular interest lies in using \CaII{43} as an ion-qubit in quantum logic experiments~\cite{97:Steane}. This ion
has a number of advantages as a qubit~\cite{03:Lucas,03:SchmidtKaler}; it is also an attractive candidate for a trapped
ion optical frequency standard~\cite{93:Plumelle,00:Boshier}. The most obvious difficulty in working with \CaII{43} is
its low natural abundance of 0.135\%. In this paper we describe the development of a photo-ionization system capable of
loading \CaII{43} into a Paul trap from a natural abundance source. (Isotopically-enriched sources are available,
although the maximum enrichment of \ish 80\%~\cite{PC:OakRidge} means that it would still be advantageous to use an
isotope-selective method for reliable loading of this isotope.)

We describe the crossed atomic and laser beam setup for excitation of the 423\nm\ 4s$^2\,\lev{1}{S}{0}\leftrightarrow$
4s4p\,\lev{1}{P}{1} transition with a Doppler width narrow compared with the isotope shifts. Spectroscopy of the atomic
beam is used to ensure the intensity of the 423\nm\ laser is below saturation (to avoid reducing the
isotope-selectivity of this step by power broadening), and also to estimate the number density of atoms in the beam.
Photo-ionization from the 4s4p\,\lev{1}{P}{1} level is achieved by a 389\nm\ photon; since we expect the
photo-ionization cross-section to be only weakly dependent on the photon energy, we use a non-stabilized diode laser
for this step. We study the photo-ionization trap loading rate as a function of the power and detuning of the 423\nm\
laser, and as a function of the power of the 389\nm\ laser. We find that the 389\nm\ laser photo-ionizes equivalently
above or below lasing threshold, and that it can be replaced by a much cheaper incoherent source if high
photo-ionization rates are not required. We estimate the absolute photo-ionization loading efficiency and compare it
with that for electron bombardment ionization in our apparatus.

The sensitivity of the loading rate as a function of the 423\nm\ detuning determines the isotope-selectivity which can
be attained. However, for very rare species with mass number $X$ a practical limitation is the rate at which
charge-exchange replaces the desired, trapped, ions \CaII{X} by the abundant \CaII{40} ions issuing from the oven. We
demonstrate loading of all the naturally occurring isotopes, including \CaII{43} and the 0.004\%-abundant \CaII{46}.
These two species are laser-cooled for the first time in the experiments reported here. The limited isotope-selectivity
attainable for these rare ions means that some form of ``purification'' of the ions loaded is
necessary~\cite{96:Alheit,01:Toyoda}; in the case of \CaII{43} we show that this is possible with no loss of the
desired species. Hence it should be possible to load arbitrarily large crystals of this important ion.

\begin{figure*}
\begin{centre}
\includegraphics[width=\twocolfig]{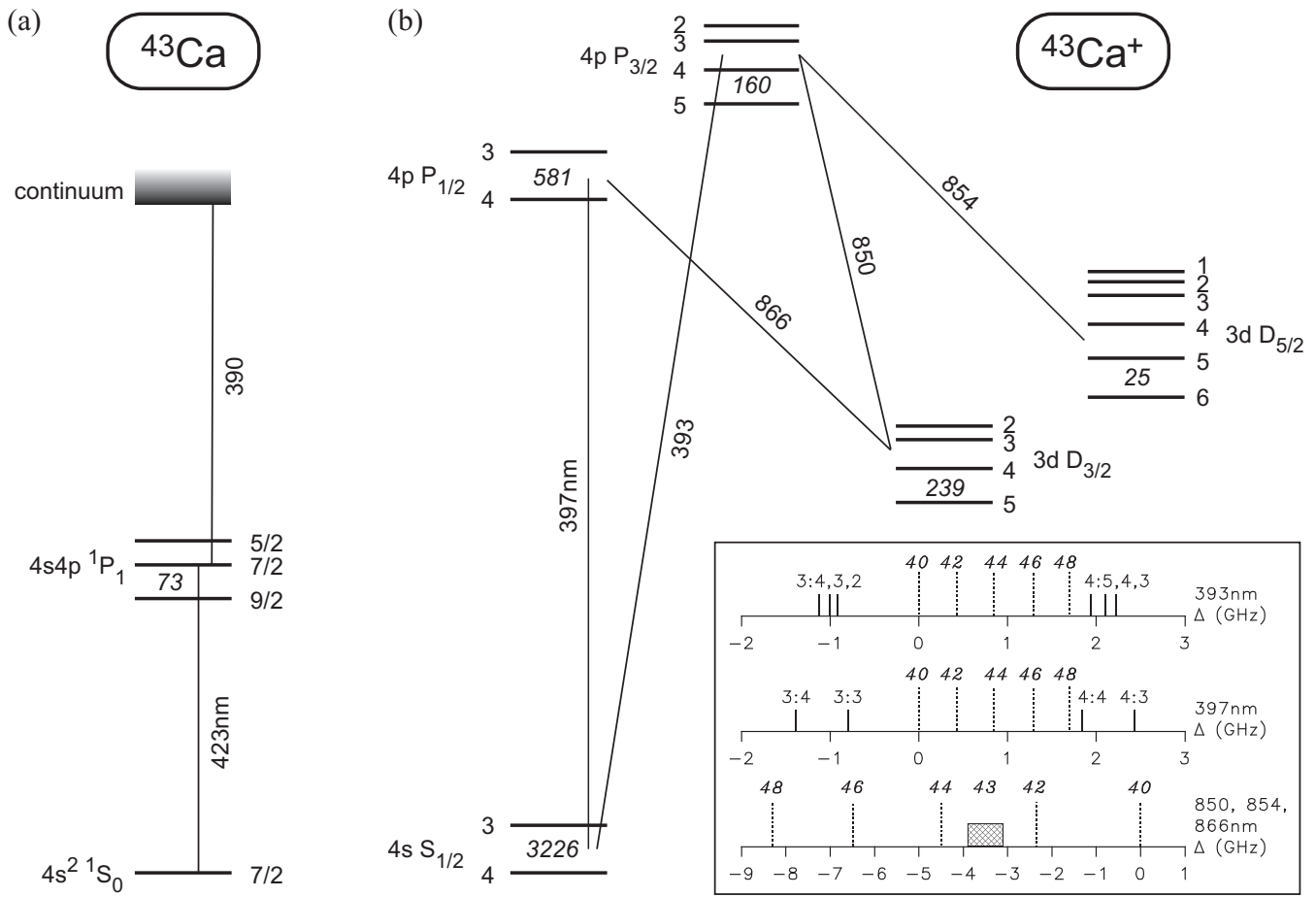}
\caption{%
Energy levels of interest in \CaI{43} (for photo-ionization) and \CaII{43} (for laser-cooling), with hyperfine levels
labelled by the total angular momentum $F$. The nuclear spin is $I=\wee{7}{2}$. The largest hyperfine splitting in each
term is indicated in MHz; the levels approximately conform to an interval
rule~\protect\cite{98a:Nortershauser,98b:Nortershauser}. Diagrams for the even isotopes are the same but with no
hyperfine structure.
(a)~\CaI{43}: the lifetime of the 4s4p\,\lev{1}{P}{1} level is 4.49(4)\ns~\cite{93:Mitroy} and the natural linewidth of
the 423\nm\ transition 35.4(3)\MHz. Photo-ionization from the 4s4p\,\lev{1}{P}{1} level requires a photon of wavelength
less than 389.8\nm.
(b)~\CaII{43}: the natural widths of the ultraviolet transitions are approximately 23\MHz, the lifetimes of the
metastable D levels about 1.2\s~\protect\cite{00:Barton}. {\em Inset:\/} Positions of the hyperfine components of the
393\nm\ and 397\nm\ transitions (solid lines, labelled with $F\sub{lower}:F\sub{upper}$) and the isotope shifts of the
even isotopes (dotted lines, labelled with the mass number), as a function of detuning $\Delta$ relative to \CaII{40}.
The isotope shifts in the infrared transitions (similar in all three transitions) are also shown, without the hyperfine
details for \CaII{43}; note the change of scale.
}%
\label{F:levels}
\end{centre}
\end{figure*}

Level diagrams for \CaI{43} and \CaII{43}, showing the energy levels and transitions relevant to these experiments, are
shown in figure~\ref{F:levels}. Isotope shifts in the six transitions shown are summarized in table~\ref{T:isodata},
together with the abundances of the six naturally occurring isotopes.

\begin{table*}
\begin{centre}
\begin{tabular}{|cc||c||c|c|} \hline
Mass & Natural & \multicolumn{3}{c|}{Isotope shifts (MHz)} \\ \cline{3-5}
number  & abundance & Ca \lev{1}{S}{0}--\lev{1}{P}{1} & Ca$^+$ S--P & Ca$^+$ D--P  \\ %
          &         & 423\nm & 397\nm\ [393\nm] & 854\nm\ [850, 866\nm] \\ \hline %
40        &  96.9\% &    0  &    0     &       0    \\ %
42        & 0.647\% &  394  &  425(6)  & $-$2350(4) \\ %
43 (c.g.) & 0.135\% &  612  &  688(17) & $-$3465(4) \\ %
44        &  2.09\% &  774  &  842(3)  & $-$4495(4) \\ %
46        & 0.004\% & 1160  & 1287(4)  & $-$6478(8) \\ %
48        & 0.187\% & 1513  & 1696(6)  & $-$8288(7) \\ \hline %
\end{tabular}
\end{centre}
\caption{Abundances of the naturally occurring calcium isotopes~\cite{02:Coplen}, and isotope shifts for Ca and Ca$^+$
transitions relevant to this work. Shifts in the 423\nm\ transition are taken from~\cite{98b:Nortershauser};
uncertainties quoted are below 1\MHz. Shifts in the ultraviolet Ca$^+$ transitions were measured
in~\cite{92:Martensson}; we quote the (more accurate) measurements for the 397\nm\ transition: the shifts at 393\nm\
are all consistent with these. Shifts in the infrared transitions are from~\cite{98a:Nortershauser}; the data given are
for the 854\nm\ transition, which are again consistent with those for the other two lines. For the odd isotope, the
shift of the centre of gravity of the hyperfine components of each transition is given; note that in the Ca$^+$ S--P
lines, the hyperfine structure is significantly
larger than the isotope shift, figure~\protect\ref{F:levels}(b).} %
\label{T:isodata}
\end{table*}

\section{Apparatus}

\subsection{Ion trap}

The apparatus consists of a linear radio-frequency (r.f., 6.2\MHz) Paul trap in an ultra-high vacuum system (background
pressure $<2\times 10^{-11}\torr$), with diode laser systems available to drive all the transitions shown in
figure~\ref{F:levels}. Details of the trap are given in~\cite{00:Barton}. Typical axial and radial frequencies used in
the present experiments are 170\kHz\ and 500\kHz\ respectively. Auxiliary d.c.\ electrodes are used to compensate stray
electric fields to the level of about 1\unit{V/m} using a photon-r.f.\ correlation technique applied to the
fluorescence from a single trapped ion~\cite{98:Berkeland}. A magnetic field of \ish 3\unit{G} is applied to prevent
optical pumping into magnetic sub-states~\cite{00:Boshier}.

\subsection{Lasers}
\label{S:lasers}

The lasers are all of the external-cavity grating-stabilized Littrow design, except the 389\nm\ laser which has no
grating stabilization. The blue and ultraviolet laser diodes are all GaN devices, the infrared ones GaAlAs. The
grating-stabilized lasers may be locked to tunable, stabilized, optical cavities to reduce linewidths and medium-term
frequency drift below 5\MHz. The laser frequencies are set within about 100\MHz\ of the ionic transitions with
wavemeters. Doppler-free saturated absorption spectroscopy in a calcium hollow cathode is used to monitor the 423\nm\
laser frequency relative to the 4s$^2\,\lev{1}{S}{0}\leftrightarrow$ 4s4p\,\lev{1}{P}{1} atomic transition. The vacuum
wavelength of the 389\nm\ free-running laser was measured using a grating spectrograph to be 388.9(1)\nm\ at 20\degC\
(with negligible dependence on drive current); above lasing threshold, the spectral width is expected to be \ltish
100\MHz, below threshold \ish 3\nm. For convenience of alignment, the 423\nm\ and 389\nm\ laser beams are both injected
into the same single-mode optical fibre to transport them to the trap; this ensures that the beams are well overlapped
in the interaction region, which is centred on the trap. The transmission of the fibre is \ish 30\% at 423\nm\ and \ish
10\% at 389\nm. The maximum intensities at the centre of these beams are \ish 50\unit{mW/mm$^2$} at 423\nm\ and \ish
5\unit{mW/mm$^2$} at 389\nm. Typical intensities of the cooling lasers at the trap are: 10\unit{mW/mm$^2$} at 397\nm;
20\unit{mW/mm$^2$} at 393\nm; 6\unit{mW/mm$^2$} at 866\nm; 200\unit{mW/mm$^2$} at 854\nm; 650\unit{mW/mm$^2$} at
850\nm. These intensities are based on the measured spot sizes for the various beams, which range from 40--300\um.

\subsection{Detection system}

Fluorescence from the interaction region at 423\nm\ and 397\nm\ acts as a diagnostic of neutral calcium and calcium
ions respectively. The trap region is imaged by a compound lens onto an aperture to reject scattered light; further
lenses re-image the trap, via a violet filter, onto a photomultiplier (PMT). A movable beamsplitter allows a portion of
the light to be directed to a charge-coupled device (CCD) camera; the camera is used to detect the presence of
non-fluorescing ions in the trap (revealed as ``gaps'' in an otherwise regular crystal), and to check whether the ions
form a cold crystal or a hot cloud. The net detection efficiency of the PMT system, including the solid angle subtended
by the lens and the quantum efficiency of the PMT, was measured to be 0.13(2)\% at 397\nm; the photon count rate
observed from a single, cold, \CaII{40} ion under conditions of near-saturation is consistent with this figure. The net
detection efficiency at 423\nm\ is $\eta=0.15(3)\%$.

The discriminated pulses from the PMT are counted directly by 10\MHz\ on-board counters in the computer which controls
the experiment; two alternately-gated counters are used to eliminate read-out delay. For the large peak photon counting
rates encountered in studies of the 423\nm\ fluorescence (up to \ish 10$^6$\unit{s$^{-1}$}), we correct the measured
PMT count rate $S'$ for the maximum counting rate $S\sub{max}=10\MHz$, assuming $S\ll S\sub{max}$, to obtain the true
count rate $S\approx S'/(1-S'/S\sub{max})$.

\subsection{Calcium beam}
\label{S:beam}

The calcium source consists of an oven made from stainless steel tube (diameter 2\mm, wall thickness 0.1\mm) filled
with granules of calcium metal, closed by crimping at each end, and with an orifice (estimated area \ish
1\unit{mm$^2$}) filed in the side facing towards the trap. The oven is heated by passing a current (between 3\unit{A}
and 6\unit{A}) along its length, which produces a beam of calcium atoms directed towards the trap. Estimated oven
temperatures are in the range 500--650\K, depending on the current used (section~\ref{S:oven423}).

The oven is situated 22(2)\mm\ from the centre of the trap. The atomic beam is collimated by the r.f.\ trap electrodes
to an effective aperture of length $l=1.7(1)\mm$ along the direction of the 423\nm/389\nm\ laser beams. However, the
detection system only images a limited length of the interaction region (inset figure~\ref{F:spectrum423}), so the
effective aperture for observation of the 423\nm\ fluorescence is reduced to $l'=0.44(4)\mm$. Since the oven orifice is
of comparable dimensions to $l,l'$ the angular divergence of the beam is somewhat larger than these distances would
otherwise imply.

\section{Spectroscopy of the atomic beam}

We describe in this section spectroscopy of the 423\nm\ transition, using the calcium atomic beam from the oven. For
optimal isotope-selectivity in the photo-ionization loading, the narrowest possible linewidth of this transition is
required. Ideally the linewidth should be small compared with the isotope shifts (table~\ref{T:isodata}), with the
lower limit being set by the 35.4(3)\MHz\ natural width. Other contributions to the homogeneous width include power
broadening (saturation), transit-time broadening and laser linewidth. Collisional broadening is entirely negligible at
the low number densities involved here. The inhomogeneous broadening (Doppler width) is determined by the oven
temperature, the collimation angle of the atomic beam and the angle between the laser beam and the atomic beam.

For orthogonal beams, the velocity distribution approximates to a Gaussian whose width compared with that in the oven
is determined by the collimation geometry and the oven orifice~\citesec{Bk:Demtroder}{10.1}. For non-orthogonal beams,
the velocity distribution broadens and becomes asymmetric, resulting in a broadening and shift when it is convolved
with the homogeneous profile.

\subsection{Alignment of the crossed beams}

\begin{figure}
\begin{centre}
\includegraphics[width=\onecolfig]{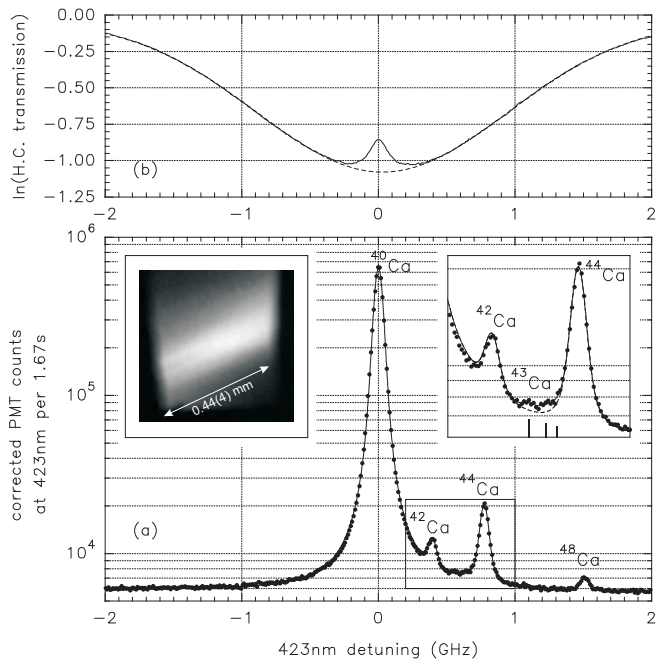}
\caption{%
(a)~Fluorescence spectrum from the atomic beam at an oven current of 5.0\unit{A} and a laser power below saturation,
$P_{423}\ish 6\uW$. The laser was locked to a tunable, stabilized, optical cavity to minimize drift during the scan;
the frequency scale was determined using a second, confocal, cavity (600\MHz\ free spectral range). The best-fit Voigt
profile is shown (solid line), where the amplitude, linear baseline, frequency offset, Lorentzian and Gaussian widths
were all floated simultaneously. The total fwhm is 76\MHz. The abundances, isotope shifts and hyperfine splittings were
given the known values~\protect\cite{98b:Nortershauser,02:Coplen}. {\em Right inset:\/} Magnification of the region
containing the three hyperfine components of \CaI{43}, whose expected positions are indicated. The dashed line shows
the best fit when the \CaI{43} components are omitted from the theoretical profile. The largest contribution to the
signal by \CaI{43} is about 22\% of the total signal above background. {\em Left inset:\/} Fluorescence at \CaI{40}
line centre from the interaction region, as observed on the CCD camera. The imaging system aperture limits the length
of the 423\nm\ beam which is observed to 0.44(4)\mm.
(b)~Simultaneous Doppler-free saturated absorption signal from the hollow cathode lamp; logarithmic transmission is
shown. The Doppler-free peak (fwhm \ish 170\MHz) is superimposed on the Doppler-broadened background (fwhm \ish
2.2\GHz, dashed curve), and is centred within 10\MHz\ of the \CaI{40} fluorescence peak. Peaks due to the other
isotopes are too small to be visible.
}%
\label{F:spectrum423}
\end{centre}
\end{figure}

To ensure orthogonal alignment of the laser and atomic beams, we minimize the width of the 423\nm\ fluorescence
spectrum as a function of the angle between the beams. The power of the 423\nm\ laser $P_{423}$ is set safely below the
saturation level for maximum sensitivity (see next section). A fluorescence spectrum with the alignment optimized is
shown in figure~\ref{F:spectrum423}, together with a simultaneously acquired Doppler-free hollow cathode spectrum. From
several such spectra we find that the offset between the \CaI{40} fluorescence peak and the hollow cathode
saturated-absorption peak is at most 10\MHz\ (after accounting for a pressure red-shift of up to 13\MHz\ in the hollow
cathode lamp due to 6\torr\ of neon buffer gas~\cite{72:Smith}). This indicates a beam misalignment of at most \ish
12\mrad.

The fitted Voigt profile shown has Lorentzian full-width at half-maximum (fwhm) 39\MHz\ and Gaussian fwhm 52\MHz. The
transit-time broadening is estimated from the measured laser spot size and the approximate atomic velocity
(section~\ref{S:oven423}) to be 2(1)\MHz. The laser linewidth estimated from the error signal when the laser is locked
to a stabilized reference cavity is 4(1)\MHz. The Gaussian width is consistent with the collimation aperture
$l'=0.44(4)\mm$ and the estimated oven temperature $T\ish 610\K$ if we model the oven as an extended source of size
\ish 1\mm.

\subsection{Saturation broadening}

\begin{figure}
\begin{centre}
\includegraphics[width=\onecolfig]{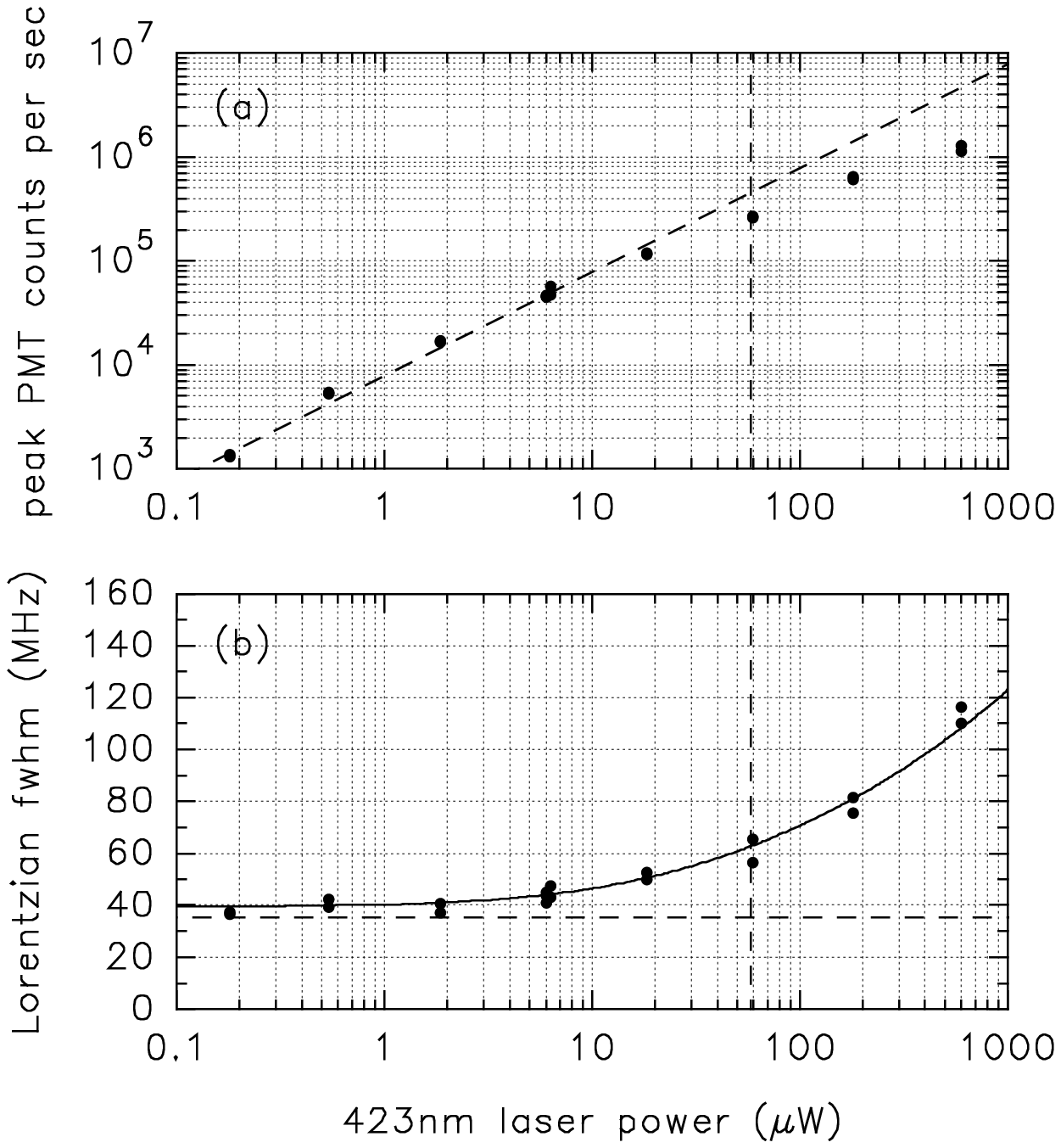}
\caption{%
Fluorescence and Lorentzian linewidth on the 423\nm\ transition, as a function of laser power $P_{423}$, obtained from
fitting 423\nm\ spectra with Voigt profiles (whose Gaussian width was fixed at 51\MHz). The oven current was
4.25\unit{A}. The vertical dashed lines indicate the power at which the intensity at the centre of a beam with 100\um\
spot size reaches $I\sub{sat}=2\pi h c A/\lambda^3=3.68\unit{mW/mm$^2$}$.
(a)~Peak (corrected) PMT count rate above background on the \CaI{40} component. A straight line ($y\propto x$) is
fitted to the data for $P_{423}<10\uW$; the onset of saturation is visible as the signal drops below this line at high
powers.
(b)~Analysed Lorentzian fwhm for the same data. The linewidth starts to increase for power above about 10\uW. The
horizontal dashed line indicates the 35\MHz\ natural linewidth. The curve shows the expected homogeneous fwhm, obtained
by averaging the atomic response over a circular Gaussian beam with 100\um\ spot size and assuming a 4\MHz\ laser
linewidth.
}%
\label{F:saturation}
\end{centre}
\end{figure}

It is important to avoid saturation broadening of the 423\nm\ transition. Since it is difficult to estimate absolute
intensities accurately, and the atomic response is averaged over the laser beam profile, we studied the saturation of
the transition empirically, by varying the laser power. Results, obtained after alignment of the crossed beams, are
shown in figure~\ref{F:saturation}. Saturation effects begin, both in the peak fluorescence and the Lorentzian
linewidth, above a laser power of about 10\uW.

The photo-ionization rate will increase as the population of the 4s4p\,\lev{1}{P}{1} state, \ie, linearly until
saturation sets in. Since the atomic beam is much larger than the laser beam, however, the total number of atoms in the
upper state will continue to increase even well above saturation, though more slowly with increasing power. The
isotope-selectivity will decrease roughly as the square of the Lorentzian linewidth, for fixed Gaussian width (the
wings of the Voigt profile are determined overwhelmingly by the Lorentzian component, for the Gaussian width involved
here). To maximize the photo-ionization rate without sacrificing isotope-selectivity, we work at powers in the range
5--15\uW\ for photo-ionization loading.

Specifically, the fractional population of the 4s4p\,\lev{1}{P}{1}\,$M_J=0$ upper state as a function of detuning
$\Delta_{423}$ at $\lambda=423\nm$ is, for linearly-polarized laser light of intensity $I$,
\begin{equation}
n\sub{4s4p}(\Delta_{423}) = \frac{\wee{3}{2} I A \beta}{(2\pi\Delta_{423})^2 + \beta^2 + 3 I A \beta} %
\label{E:pop4s4p}
\end{equation}
where $A$ is the Einstein coefficient for the transition ($A=2.23\times 10^8\unit{s$^{-1}$}$; the weak decay route to
4s3d\,\lev{1}{D}{2} may be neglected since the calculated lifetime against this decay~\cite{93:Mitroy} is long compared
to the transit time across the laser beam), $\beta=(2\pi\Gamma\sub{L}+A)/2$ with $\Gamma\sub{L}$ the laser linewidth,
the intensity is measured in units of a saturation intensity $I\sub{sat}=2\pi h c A/\lambda^3$ and $\Delta_{423},
\Gamma\sub{L}$ are in units of Hz. This expression must be integrated over the spatial profile of the laser beam to
obtain the expected signal from the atomic beam; the resulting homogeneous fwhm, assuming a circular Gaussian beam and
a laser linewidth $\Gamma\sub{L}=4\MHz$, is plotted in figure~\ref{F:saturation}(b).

\subsection{Number density and temperature in the atomic beam}
\label{S:oven423}

We can use the peak \CaI{40} fluorescence at 423\nm\ to deduce the number density of calcium atoms in the atomic beam,
using the known line strength and the measured photon collection efficiency. This in turn permits an estimate of the
absolute photo-ionization loading efficiency. By measuring the number density as a function of oven current we can also
compare the efficiency of photo-ionization trap loading with the electron bombardment method (which have similar
loading rates at very different oven currents).

\begin{figure}
\begin{centre}
\includegraphics[width=\onecolfig]{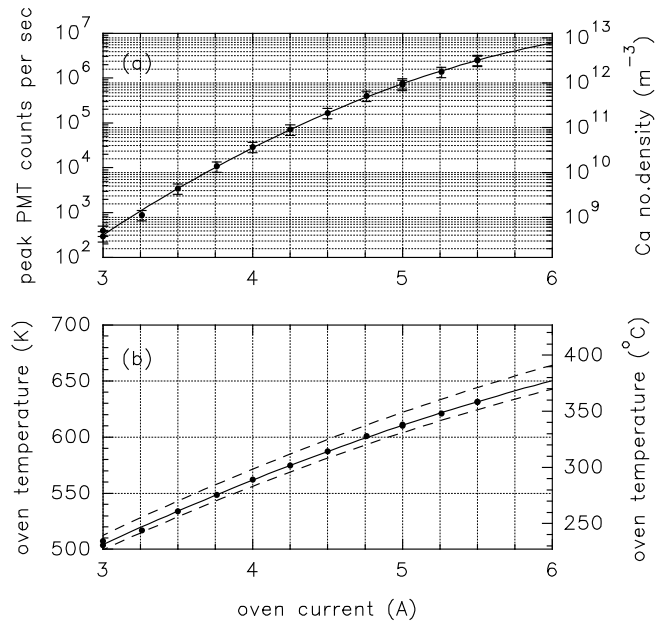}
\caption{%
Calcium number density and oven temperature as a function of oven current.
(a)~Corrected photon count rate above background on the \CaI{40} peak obtained from fitting 423\nm\ spectra with Voigt
profiles, where the Lorentzian width was fixed at 41\MHz\ from a spectrum at 4.25\unit{A} oven current. The right-hand
ordinate gives the calcium number density in the interaction region deduced from this fluorescence rate as described in
the text, and the error bars refer to the number density. A quadratic regression fit is shown (solid line). At
3\unit{A}, the signal was so small that the line shape was poorly determined and the Gaussian width was also fixed for
the fit.
(b)~Estimated oven temperature based on the number density in (a), with quadratic regression (solid line). The dashed
lines indicate the temperature range implied by the 50\% uncertainty assigned to the oven orifice area
$\sigma=1.0(5)\unit{mm$^2$}$.
}%
\label{F:oven423}
\end{centre}
\end{figure}

The peak fluorescence as a function of oven current is shown in figure~\ref{F:oven423}(a). The laser power was set
below saturation and under these conditions the peak corrected photon count rate $S$ is related to the number density
$N$ of atoms in the ground state by
\[ S = \frac{3\lambda^3 A}{8\pi h c} \, P_{423} l' \eta r \cdot {\cal V}(0) \cdot a_{40} N \]
where $a_{40}=96.9\%$ is the \CaI{40} abundance, $l'=0.44(4)\mm$ the length of the interaction region observed through
the imaging aperture (measured along the laser beam direction), $\eta=0.15(3)\%$ the net collection efficiency of the
detection system at $\lambda=423\nm$, $r=1.43(2)$ a geometrical factor arising from the angular distribution of the
fluorescence and ${\cal V}(0)$ the value of the fitted Voigt profile at line centre (which is normalized such that
$\int_{0}^{\infty} {\cal V}(\nu-\nu_0) \diff{\nu}=1$). The fitted values of ${\cal V}(0)$ show a slight decrease with
increasing oven current, due to increasing inhomogeneous width as the temperature rises, but this is no greater than
the \ish 10\% fitting uncertainty, and for the purpose of figure~\ref{F:oven423}(a) we use the average value
$\overline{{\cal V}(0)}=9.9(1.0)\unit{GHz$^{-1}$}$ so that the number density $N[\unit{m$^{-3}$}] = 1.3(3)\times 10^6
S[\unit{s$^{-1}$}]$ and it can be plotted on the same ordinate.

For a small interaction region on axis of the atomic beam, the number density $N$ in the interaction region is related
to that in the oven $N_0$ by
\[ N = \frac{N_0 \sigma}{4\pi d^2} \]
where $\sigma$ is the area of the oven orifice and $d$ the distance from the orifice to the interaction
region~\citesec{Bk:Scoles}{4.2}. In our apparatus, $d=22(2)\mm$ but $\sigma$ is not accurately known. However, since
number density depends strongly on temperature $T$, we can make a reasonable estimate of $T$ by assuming that
$\sigma=1.0(5)\unit{mm$^2$}$, the value suggested by our profile analysis. We then find a temperature which satisfies
$p=N_0 k T$ and the known vapour pressure curve $p=p(T)$ simultaneously~\cite{Bk:Barin}. This is shown in
figure~\ref{F:oven423}(b), and ranges between about 500\K\ and 650\K. The temperature is useful for estimating the mean
velocity along the beam $\overline{v}=\sqrt{\pi k T / 2 M}$ for atoms of mass $M$: $\overline{v}=433(4)\unit{m/s}$ for
\CaI{40} at $T=575(10)\K$.

\subsection{Conclusion}

With the 423\nm\ beam well aligned orthogonal to the atomic beam, and the laser power set below saturation, we observe
a linewidth of 76\MHz\ (fwhm), made up of Lorentzian and Gaussian contributions of about 39\MHz\ and 52\MHz\
respectively. This compares favourably with the isotope shifts, table~\ref{T:isodata}. We do not use these data to
predict relative loading probabilities for the different isotopes, however, since the collimation angle of the atomic
beam for loading the trap is greater than that relevant to the 423\nm\ fluorescence ($l>l'$) and the Gaussian
contribution to the linewidth will be correspondingly larger. A quantitative estimate of this is difficult since the
angular divergence depends on the sizes of the oven orifice and the trap capture region, neither of which is accurately
known. Instead, we investigate the actual photo-ionization loading rate as a function of 423\nm\ detuning, as described
in the following section.

\section{Photo-ionization loading studies}

\subsection{Expected form of photo-ionization cross-section}
\label{S:Xsection}

Calculations~\cite{PC:vanderHart} of the cross-section $Q$ for photo-ionization from the 4s4p state suggest that, at
389\nm, $Q$ lies in the range 60--280\unit{megabarns} ($1\unit{megabarn} = 10^{-22}\unit{m$^2$}$). The large range
results from the existence of an auto-ionizing resonance close to the ionization threshold, whose wavelength is
uncertain to \ish 5\nm. Since the energy of the resonance is not well-known, we do not attempt to choose a particular
laser wavelength, beyond ensuring that it is below the threshold required for ionization. The width of the resonance is
certainly large compared to the laser linewidth, so there is no need for the laser to be carefully
frequency-stabilized.

The electric fields present in a Paul trap can allow ionization to occur even for excitation slightly below the
continuum limit~\cite{01:Gulde}. For the typical velocity estimated above, atoms cross the photo-ionization laser beams
in \ish 0.2\us\ and all atoms are thus able to sample the extrema of the r.f.\ field (period 0.16\us). The peak field
varies across the trap, but is at most \ish 10$^5$\unit{V/m}. A rough estimate~\cite{Bk:Kuhn} of the effect of such a
field on the ionization threshold may be made by assuming a hydrogenic potential $V(x)=-e/4\pi\epsilon_0 x$ near the
continuum and equating this to the potential ($-eEx$) of the electron in the external field $E$; the result is that the
ionization limit could be shifted at the furthest to a wavelength of 391.3\nm. However, no significant enhancement of
the photo-ionization cross-section is expected by using wavelengths closer to this shifted ionization limit, even for
monochromatic light, because the oscillatory electric field blurs out the discrete energy levels of the atom near the
continuum and the average transition probability is no higher than that within the continuum~\citesec{Bk:Cowan}{18.6}.

\subsection{Photo-ionization loading rate measurements}

\begin{figure}
\begin{centre}
\includegraphics[width=\onecolfig]{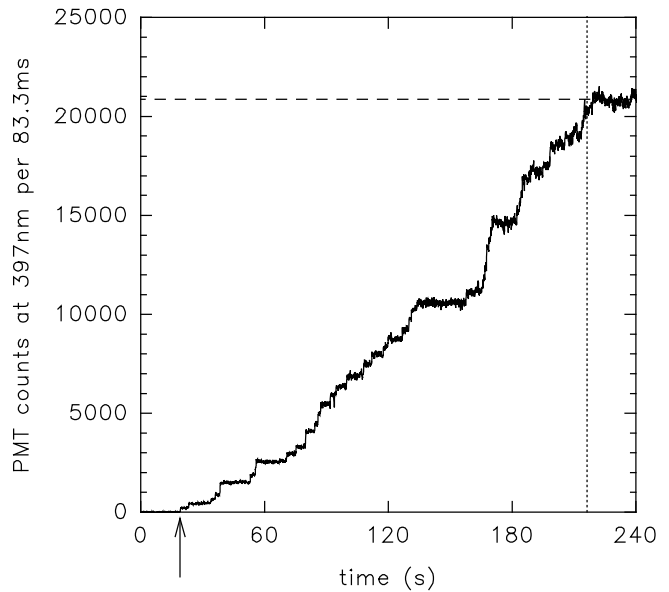}
\caption{%
Typical photo-ionization loading fluorescence curve. The 423\nm\ and 389\nm\ photo-ionizing lasers are switched on at
$t=0$ and off at $t=216\s$ (dotted line). The 397\nm\ and 866\nm\ cooling lasers illuminate the ions continuously.
Background counts due to scattered 397\nm\ light and (during the loading) 423\nm\ fluorescence and 389\nm\ light have
been subtracted from the PMT signal. The step in 397\nm\ fluorescence when the first cold ion appears is arrowed. The
large steps in the fluorescence are typical and may result from the details of the cooling dynamics as the number of
ions in the trap increases. The dashed line indicates the final fluorescence level, which corresponds to 36(3) ions;
the average loading rate of 0.17(3)\unit{ion/s} is obtained by dividing this by the time for which the photo-ionizing
lasers are switched on. The conditions were: oven current 4.25\unit{A}, $P_{423}\ish 6\uW, \Delta_{423}=0(5)\MHz,
P_{389}=0.93(9)\uW, \Delta_{397}=-68(10)\MHz$.
}%
\label{F:PIload}
\end{centre}
\end{figure}

\begin{figure}
\begin{centre}
\includegraphics[width=\onecolfig]{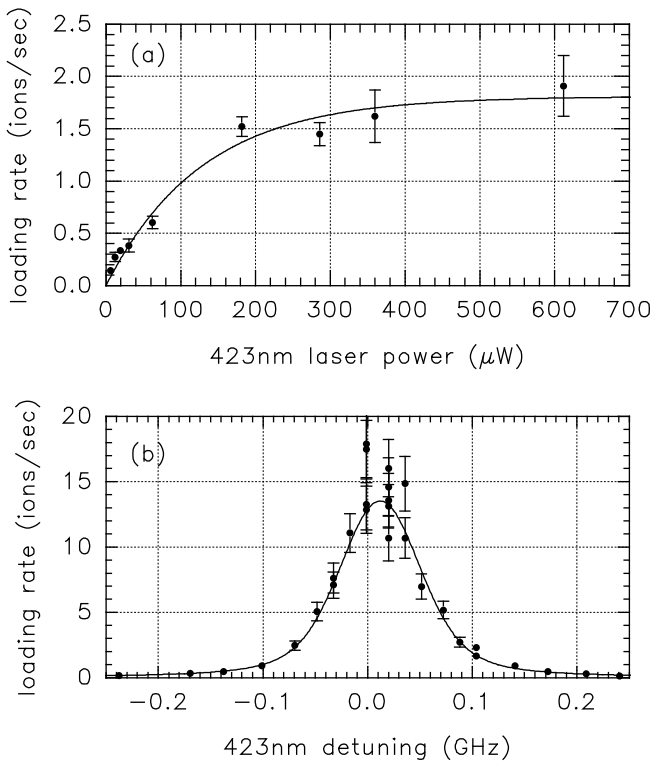}
\caption{%
(a)~Photo-ionization loading rate as a function of 423\nm\ laser power, with: oven current 4.25\unit{A},
$\Delta_{423}=0(10)\MHz, P_{389}=0.93(9)\uW, \Delta_{397}=-80(20)\MHz$. The loading rate starts to saturate for
$P_{423}\gtish 10\uW$. The curve is drawn to guide the eye.
(b)~Loading rate versus 423\nm\ detuning, with: oven current 4.25\unit{A}, $P_{423}\ish 6\uW, P_{389}=80(8)\uW,
\Delta_{397}=-97(10)\MHz$. A Voigt profile (solid line) is fitted to the data; the frequency offset, amplitude and
Gaussian width were floated whilst the Lorentzian width was fixed at 39\MHz\ from the fit in
figure~\protect\ref{F:spectrum423}. The fwhm of the fitted curve is 94\MHz. The frequency origin is taken from the
fitted offset of a 423\nm\ fluorescence spectrum taken during the same run; the fitted profile here shows a blue-shift
of 12\MHz\ which probably results from the slightly different collimation geometry relevant to loading.
}%
\label{F:PD423}
\end{centre}
\end{figure}

A typical photo-ionization loading curve is shown in figure~\ref{F:PIload}, where the fluorescence at 397\nm\ from cold
\CaII{40} ions in the trap is plotted as a function of time after the photo-ionizing lasers are switched on. As the
size of the ion crystal grows, the fluorescence per ion increases due to the effect of r.f.\ micromotion for ions lying
off-axis, because they spend time closer to resonance with the red-detuned 397\nm\ cooling laser. However, for the
range of crystal sizes in these experiments, the fluorescence per ion is roughly constant and we use the final
fluorescence level as a measure of the total number of ions loaded. (The fluorescence was calibrated after a typical
load by reducing the trap strength until the ions could be counted on the camera.) For each load, we wait a few seconds
after the photo-ionizing lasers are switched off, to allow any hot ions to be cooled and to join the crystal, check
that the ions are still crystalline using the camera, and then take the final number of ions $m$ divided by the
duration $\tau$ of exposure to the photo-ionizing lasers as our measure of the loading rate $R$. We find that an error
$\sqrt{m}/\tau$ is broadly consistent with variations in the loading rate taken under nominally identical conditions
(see for example figure~\ref{F:PD423}).

We have studied the loading rate as a function of the power $P_{423}$ and detuning $\Delta_{423}$ of the 423\nm\ laser,
and of the power $P_{389}$ of the 389\nm\ laser. This was done at fixed oven current, to keep the atomic number density
as constant as possible. For each measurement, the 397\nm\ fluorescence was allowed to increase to a level representing
about 40 trapped ions, whereupon the photo-ionizing lasers were blocked. In practice, the number of ions loaded varied
between 22(2) and 63(5); the length of photo-ionizing exposure was between 2\s\ and 600\s.

Figure~\ref{F:PD423} shows the dependence of loading rate on 423\nm\ power and detuning. Note that the loading rate
could in principle saturate at a different power level to the 423\nm\ fluorescence studied above, since
photo-ionization is limited by the width of the 389\nm\ beam (the lifetime of the 4s4p\,\lev{1}{P}{1} level,
4.49(4)\ns, is much shorter than the time an atom spends in the laser beams, \ish 0.2\us) whereas the fluorescence
collected is limited by the imaging aperture. Nevertheless, we observe that the loading rate also starts to saturate
above a power of $P_{423}\ish 10\uW$ due to saturation of the upper state population. Given the power available in the
423\nm\ laser, the maximum loading rate could be made perhaps an order of magnitude larger by increasing the spot size
at the trap to 1\unit{mm}; spot sizes much larger than this, however, would start to reduce the isotope-selectivity
because of the increasing atomic beam divergence angle.

The loading rate as a function of $\Delta_{423}$ is of critical importance for isotope-selective loading. We expect the
dependence to be modelled by a Voigt profile with similar Lorentzian width to that of the 423\nm\ fluorescence
spectrum, but different Gaussian width because of the different atomic beam collimation angles for fluorescence
detection and ion loading. The Lorentzian and Gaussian widths are not well-determined if they are both floated, due to
the noise in the data, so we fix the Lorentzian width at 39\MHz, from the fluorescence spectrum in
figure~\ref{F:spectrum423}, and find a fitted Gaussian width of 71\MHz. This Gaussian width is consistent with a beam
collimation angle defined by the electrodes and an oven orifice of size \ish 1\mm. The consequent isotope-selectivity
expected is discussed in section~\ref{S:isosel}.

\begin{figure}
\begin{centre}
\includegraphics[width=\onecolfig]{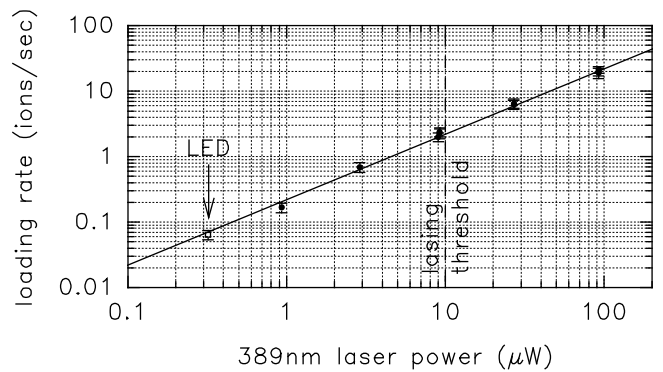}
\caption{%
Photo-ionization loading rate as a function of 389\nm\ laser power, with: oven current 4.25\unit{A}, $P_{423}\ish 6\uW,
\Delta_{423}=0(5)\MHz, \Delta_{397}=-68(10)\MHz$. No significant difference in efficiency is observed even when the
laser is well below lasing threshold (vertical dashed line), indicating that the photo-ionization cross-section is
independent of the spectral width of the laser. A straight line ($y\propto x$) is fitted to the data. The single
measurement with the ultraviolet LED in place of the 389\nm\ laser is also included (see
section~\protect\ref{S:uvLED}).
}%
\label{F:P389}
\end{centre}
\end{figure}

In contrast to the 423\nm\ bound-bound transition, no saturation effect is expected for the 389\nm\ bound-free ionizing
transition, and this is verified for the 389\nm\ powers we can attain (figure~\ref{F:P389}): the dependence of the
loading rate on $P_{389}$ is linear over two orders of magnitude. Significantly, there is no change in efficiency
whether the 389\nm\ laser is running above or below lasing threshold, despite the fact that the spectral width is
expected to be about five orders of magnitude larger (\ish 3\nm) below threshold. This implies a weak dependence on
wavelength of the photo-ionization cross-section in this region.

\subsection{Absolute photo-ionization efficiency}

We can deduce a lower limit for the photo-ionization cross-section from the loading rate under conditions of
near-saturation on the 423\nm\ transition, when we expect that nearly half the atoms are in the 4s4p\,\lev{1}{P}{1}
upper level (at least across that part of the 423\nm\ beam which overlaps the 389\nm\ beam). From the definition of the
photo-ionization cross-section $Q$, the number of atoms ionized per unit time is
\[ R = Q l \cdot \wee{1}{2} a_{40} N \cdot P_{389} / \varepsilon \]
where $l=1.7(1)\mm$ is the length of the interaction region defined by the trap electrodes, $(\wee{1}{2} a_{40} N)$ the
number density of \CaI{40} atoms in the 4s4p\,\lev{1}{P}{1} level, $P_{389}$ the 389\nm\ laser power, and $\varepsilon$
the 389\nm\ photon energy.

To minimize uncertainty in the number density due to variations in oven conditions, $N$ was determined shortly before
measuring the loading rate and found to be $1.7(4)\times 10^{11}\unit{m$^{-3}$}$. At maximum 423\nm\ power, the
estimated intensity at the centre of the laser beam was $14(4)I\sub{sat}$ which, for $|\Delta_{423}| < 10\MHz$, gives
$n\sub{4s4p}>0.48(1)$ within the half-intensity radius of the beam (equation~\ref{E:pop4s4p}), justifying the
assumption of saturation. The loading rate was measured to be $R=4.9(9)\unit{ion/s}$ with $P_{389}=1.1(1)\uW$, giving a
cross-section $Q=170(60)\unit{megabarns}$. This is a lower limit since it assumes that every ion produced is trapped
and that the capture region of the trap extends across the full width of the atomic beam as far as the r.f.\
electrodes, but it lies in the anticipated range (section~\ref{S:Xsection}).%

Alternatively, the loading rate can be expressed in terms of the probability $q$ per atom of being ionized (and
trapped) as it crosses the interaction region:
\[ R = q \cdot \wee{1}{2} a_{40} N \cdot l w^2 / t \]
where $w^2$ is the cross-sectional area of the 389\nm\ beam (which limits the extent of the interaction region), and
$t\approx w/v$ is the time an atom of velocity $v$ spends in the interaction region. For the conditions above and the
laser beam size $w\approx 0.1\mm$, we have $t\approx 0.2\us$ and find $q\approx 8\times 10^{-7}$. The maximum $q$ we
observed (with maximum powers in both the photo-ionizing lasers) was $q\approx 4\times 10^{-5}$, and by eliminating the
optical fibre and using the maximum available 389\nm\ power \ish 2\mW, we would in principle be able to obtain a
maximum $q\approx 10^{-3}$, \ie, one in 1000 atoms crossing the interaction region would be loaded.

\subsection{Loading using an ultraviolet LED}
\label{S:uvLED}

The observation that the 389\nm\ laser is able to photo-ionize atoms just as efficiently below lasing threshold led us
to experiment with an alternative light source for the second step of the photo-ionization, an ultraviolet
light-emitting diode (LED). The device tested was a Nichia NSHU590 diode, with nominal output power 500\uW, peak
wavelength 375\nm\ and spectral width 12\nm. The chief advantages of this device are that it is about three orders of
magnitude cheaper than a wavelength-selected ultraviolet laser diode, and does not require temperature and current
stabilization. Measurements with a grating spectrograph indicated that the peak wavelength was in fact between 375\nm\
and 390\nm, depending which part of the emission region was focussed on the spectrograph slit, and the spectral width
was \ish 25\nm; however at least half of the spectral intensity lies in the range $\lambda<389.8\nm$ useful for
photo-ionization. The main drawback of the LED is its extended emission region: it is difficult to focus a large
fraction of the power into a region smaller than a few millimetres, unlike a laser beam.

When we first tested the LED in place of the 389\nm\ laser we were surprised to find that it was capable of loading
ions into the trap while the 423\nm\ laser was blocked and the oven was switched off! A small number of \CaII{40} ions
were loaded, together with several ``dark'' ions; at the same time we noticed that the stray d.c.\ electric field in
the trap changed by \ish 10\%, drifting gradually back to its former value over a period of an hour or so. Our
interpretation is that the large (\ish 5\mm) patch of light from the LED was irradiating the r.f.\ electrodes, ablating
and photo-ionizing calcium and other ions from the surface where they had previously been deposited by the atomic beam.
Photo-electric emission could be responsible for charging insulating patches on the electrodes, causing the stray field
to change temporarily before the charge leaks away (the photo-electric cut-off wavelength for calcium is 433\nm).
Surface ablation is a well-known method of loading ion traps~\cite{02:Blinov}, though it is perhaps surprising at the
low light intensities involved here.

To avoid this problem, we imaged the LED onto a 200\um\ pinhole, which was re-imaged at approximately 1:1 magnification
into the trap region, and overlapped with the 423\nm\ beam. This eliminated loading of non-\CaII{40} ions and effects
on the stray field, but only 0.32(3)\uW of ultraviolet power was then available at the trap. The loading rate under
these conditions, 0.06(1)\unit{ion/s}, is plotted in figure~\ref{F:P389} and is similar to that expected with
comparable 389\nm\ laser power. This rate is not sufficient for loading the rarer isotopes, where it is necessary to
beat losses due to charge-exchange (section~\ref{S:exchange}), but is perfectly adequate for loading small crystals of
\CaII{40}. We note that high-power (\ish 150\mW) ultraviolet LEDs are also available, though their cost approaches that
of a laser diode.

\section{Isotope-selective loading}

\subsection{Expected selectivity}
\label{S:isosel}

\begin{figure}
\begin{centre}
\includegraphics[width=\onecolfig]{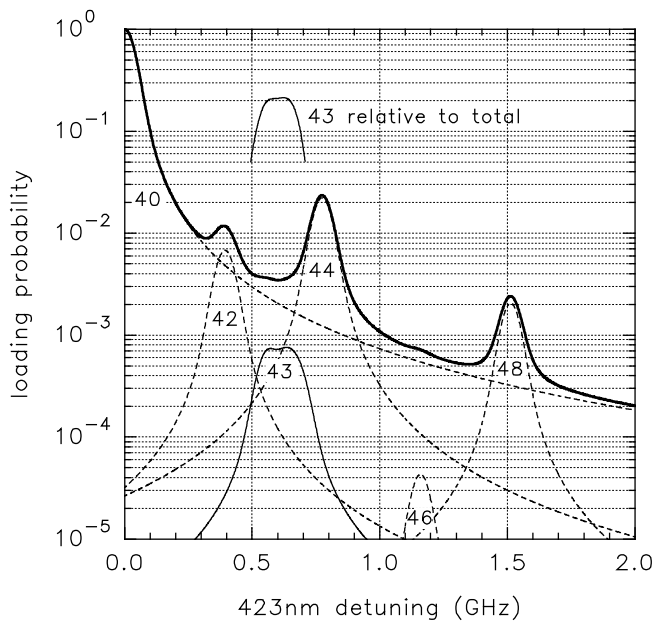}
\caption{%
Probability of loading different calcium isotopes as a function of detuning of the 423\nm\ laser, normalized to 1 for
\CaII{40} at zero detuning. The curves are generated using Voigt profiles with the same width parameters as the fit in
figure~\protect\ref{F:PD423}(b) (except that the Gaussian width is adjusted to take account of the different isotopic
masses). The dashed curves for the even isotopes, and the solid curve for \CaII{43} (the three hyperfine components
have been summed), are labelled by the mass numbers. The thick solid curve is the sum of the contributions from all the
isotopes. The detached solid curve gives the probability of loading \CaII{43} relative to the total loading probability
at the same detuning, plotted in the region where this quantity is $>0.05$.
}%
\label{F:isosel}
\end{centre}
\end{figure}

From the data in figure~\ref{F:PD423}(b) we expect good isotope-selectivity, because the net fwhm, 94\MHz, is small
compared with the isotope shifts in the 423\nm\ transition. This discrimination has to be set against the very low
abundance of some isotopes. In figure~\ref{F:isosel} we plot the calculated probability of loading the different
isotopes as a function of detuning $\Delta_{423}$, based on the fitted Voigt profile in figure~\ref{F:PD423}(b). We
regard the loading as ``isotope-selective'' for those isotopes where there exists a detuning at which one is more
likely to load the desired isotope than any other; it can be seen from figure~\ref{F:isosel} that this is the case for
all isotopes except \CaII{43} and \CaII{46}. The most demanding case is \CaII{46}, where the maximum achievable loading
probability, relative to the total loading probability, is only 6.0\%. The relative probability of loading \CaII{43} is
also plotted in the figure: it reaches a maximum of 21\% at a detuning of 0.623\GHz. This compares with a theoretical
maximum of 33\% which would be possible with a Lorentzian width equal to the natural linewidth and negligible Gaussian
width; thus not a great deal would be gained by efforts to collimate further the atomic beam.

The detunings of the Doppler-cooling lasers also affect the selectivity, due to the isotope shifts in the ionic
transitions (table~\ref{T:isodata}). For example, to cool \CaII{42} efficiently, the 397\nm\ laser needs to be detuned
\ish 350\MHz\ to the blue of the \CaII{40} transition, which will tend to heat \CaII{40} ions and expel them from the
trap, thus enhancing the selectivity.

\subsection{Charge-exchange loss}
\label{S:exchange}

It is possible for an ion \CaII{X} which has been loaded into the trap to undergo charge-exchange with a neutral atom
in the atomic beam; since the overwhelming probability is that the neutral atom is \CaI{40} the process is usually:
\[ \CaII{X} + \CaI{40} \rightarrow \CaI{X} + \CaII{40} \]
and the original ion in the trap is replaced by \CaII{40}. The charge-exchange rate is proportional to the number
density in the atomic beam. This process is a limitation for loading more than one ion of a rare isotope: if the loss
rate through charge-exchange exceeds the loading rate, then rare ions already trapped are replaced by \CaII{40} faster
than subsequent ones can be loaded. If the charge-exchange rate is sufficiently fast, then rare ions will not even be
cooled before they are replaced, and the trap will fill up with \CaII{40} before any fluorescence signal from the rare
ions is seen.

In practice we find that, providing one is not attempting to load large crystals, charge-exchange is only a limitation
for the rarest isotope \CaII{46}. For the typical loading parameters used for loading \CaII{43}, for example, the
loading rate is \ish 1\unit{ion/s} while the lifetime against loss due to charge-exchange is \ish 1\unit{min}.
\CaII{46} is about thirty times less abundant, so we expect the loss rate to approach the loading rate under the same
conditions.

We note that the charge-exchange process was used by Mortensen \etal~\cite{02:Mortensen} to demonstrate
isotope-selective loading, but that they were not attempting to retain and to laser-cool the different isotopes;
instead they relied on charge-exchange to replace every isotope by \CaII{40} and used the resulting \CaII{40}
fluorescence as a measure of the original load size.

\subsection{Even isotopes}

Loading one of the even isotopes consists of choosing the 423\nm\ detuning for maximum selectivity and setting the
appropriate detunings of the 397\nm\ and 866\nm\ lasers used for Doppler-cooling. After loading, we can check for the
presence of other isotopes by looking for fluorescence using different detunings of the cooling lasers (the lasers are
blocked while the detunings are changed, to avoid possible heating). Phase-sensitive detection with a chopped 866\nm\
beam can be used to detect very small fluorescence signals above scattered 397\nm\ background light.

As expected from figure~\ref{F:isosel} we can load small, pure crystals of \CaII{44} and \CaII{48} (and of course
\CaII{40}) without difficulty. Predicted peak loading probabilities of \CaII{44} and \CaII{48} are 94\% and 85\%
respectively. Practical loading rates can be achieved in spite of the small abundances by using maximum 389\nm\ power
(\ish 100\uW) and by increasing the oven current. We concentrate here on the more challenging cases of \CaII{42} (which
lies closest to \CaII{40}) and \CaII{46} (which is extremely rare).

The maximum expected probability of loading \CaII{42} is 58\%, at a detuning $\Delta_{423}=0.394\GHz$, with \CaII{40}
by far the dominant impurity. However, when we loaded using this detuning, we obtained a crystal consisting of \ish 19
\CaII{42} ions and only \ish 5 \CaII{40} ions, \ie, a loading efficiency of about 80\%. We attribute this to the
mechanism mentioned above, that the 397\nm\ laser was blue-detuned from the \CaII{40} transition by about 350\MHz, thus
tending to heat this species and expel it from the trap; the \CaII{40} ions are only cooled sympathetically by the
\CaII{42}~\cite{99:Bowe}. This mixed-species crystal also illustrates the typical charge-exchange rates: after \ish
20\unit{min}, with the photo-ionizing lasers blocked but the oven left on at 4.25\unit{A}, the number of \CaII{40} ions
had increased to \ish 18 and there was only a very low signal visible at the \CaII{42} detunings; after a further
30\unit{min}, there was no detectable signal from \CaII{42}. This indicates that the lifetime (per ion) against loss by
charge-exchange is \ish 17\unit{min}, for a number density in the atomic beam of $9(2)\times 10^{10}\unit{m$^{-3}$}$.
The limited isotope-selectivity achievable could be overcome by selective heating methods~\cite{96:Alheit,01:Toyoda} to
obtain a pure crystal of \CaII{42}.

\begin{figure}
\begin{centre}
\includegraphics[width=\onecolfig]{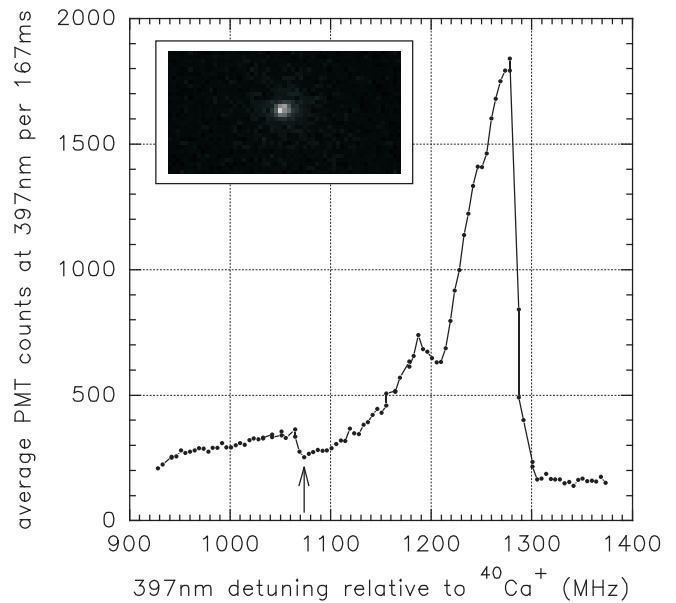}
\caption{%
Fluoresence signal from a single \CaII{46} ion, trapped in the company of several ``dark'' ions (probably \CaII{40}),
as a function of 397\nm\ detuning relative to \CaII{40}. Three scans were averaged to produce the plot. The
high-frequency ``edge'' in the fluorescence curve, characteristic of scanning past the resonant frequency, has been
assigned the known isotope shift of 1287\MHz. The first dip in the fluorescence (arrowed) is the point at which the
ions crystallize and the single \CaII{46} ion becomes visible on the camera ({\em inset}). The second dip, at \ish
1200\MHz, is probably due to a two-photon dark resonance between the 866\nm\ and 397\nm\
transitions~\protect\cite{92:Siemers}. The crystallization dip, and the movement of the single \CaII{46} ion between
discrete positions in the trap after heating and re-crystallization, indicate the presence of other ions.
}%
\label{F:ca46}
\end{centre}
\end{figure}

The 0.004\%-abundant \CaII{46} is the most difficult isotope to load, with an expected peak loading probability of
6.0\% (neglecting enhancement by heating as observed for \CaII{42}). To achieve the maximum photo-ionization rate
without excessive saturation-broadening of the 423\nm\ transition, we set $P_{423}\ish 180\uW$; we used maximum 389\nm\
power (\ish 100\uW) and increased the oven current to 5.0\unit{A}. Using the measured loading rates for \CaI{40} and
the number density data we expect a \CaII{46} loading rate of \ish 0.1\unit{ion/s}, with a sacrifice of a factor \ish 3
in selectivity because of the high 423\nm\ power. Under these conditions, a fluorescence signal was detected after \ish
17\unit{s} of exposure to the photo-ionizing lasers. The oven was switched off immediately to prevent charge-exchange.
The ions could not be made to crystallize, however, indicating that too many impurity ions were present to be cooled
sympathetically by the \CaII{46}. We therefore applied a ``tickle'' voltage to one of the trap end-caps, close to the
axial resonance frequency, which is less likely to expel the directly-cooled \CaII{46} than other isotopes.
Crystallization was then observed, figure~\ref{F:ca46}, and camera images implied that the crystal consisted of a
single \CaII{46} ion and several impurity ions (most probably \CaII{40}). The observed isotope shift in the 866\nm\
transition was $-6.4(2)\GHz$; although this particular shift has not been measured before, it is in agreement with that
expected (table~\ref{T:isodata}) and thus confirms the identity of the ion.

\subsection{Odd isotope}

The laser requirements for Doppler-cooling \CaII{43} are more demanding than for the even isotopes because of the
hyperfine structure, figure~\ref{F:levels}(b). In particular, a separate repumper is necessary on one of the
ultraviolet transitions because of the large (3.2\GHz) hyperfine splitting between the \hfslev{S}{1/2}{3} and
\hfslev{S}{1/2}{4} levels (where the superscript denotes $F$).

We choose the 393\nm\ $\hfslev{S}{1/2}{4}\leftrightarrow\hfslev{P}{3/2}{5}$ transition for cooling, both because it is
an intrinsically strong $F\leftrightarrow F+1$ transition and (thanks to the inverted hyperfine structure) it is
blue-detuned from the $\lev{}{S}{1/2}\leftrightarrow\lev{}{P}{3/2}$ transitions in {\em all\/} the even isotopes (inset
figure~\ref{F:levels}). This second fact means that any other isotopes that are loaded will tend to be heated.
Furthermore, this transition is nearly-cycling (apart from the \ish 6\% decay route to the D levels and off-resonant
driving of the 160\MHz-detuned $\hfslev{S}{1/2}{4}\leftrightarrow\hfslev{P}{3/2}{4}$ component), so that the second
ultraviolet laser needed for repumping out of the \hfslev{S}{1/2}{3} level can still repump efficiently at low
intensities: this is an advantage since it is necessarily red-detuned relative to the even isotopes and can thus cool
them. We choose to use the 397\nm\ $\hfslev{S}{1/2}{3}\leftrightarrow\hfslev{P}{1/2}{4}$ transition for this repumping.

Repumping from the \lev{}{D}{3/2} and \lev{}{D}{5/2} manifolds is accomplished with the 850\nm\ and 854\nm\ lasers
respectively: the high intensities available in these beams mean that they can repump adequately in spite of the
hyperfine structure which spans \ish 400\MHz\ in each transition.

To load we set the appropriate detunings of these four lasers relative to \CaII{40}, and set $\Delta_{423}\approx
600\MHz$. The 423\nm\ power is typically \ish 10\uW, to avoid saturation-broadening, and the 389\nm\ power maximized
(\ish 100\uW) for fastest loading. Under these conditions and with an oven current of 5.0\unit{A}, the loading rate is
\ish 1\unit{ion/s} and the lifetime against loss through charge-exchange \ish 1\unit{min}. This gives time to detect a
single ion of \CaII{43} and switch the oven off before it is replaced by one of \CaII{40}. With this procedure, we can
reliably load single ions of \CaII{43} and only rarely find that we have loaded \CaII{40} as well; this indicates that
we achieve significant enhancement over the estimated maximum loading probability of only 21\%, presumably because
\CaII{40} ions are heated by the 393\nm\ laser and expelled from the trap. The maximum fluorescence rate detected from
a single \CaII{43} ion was 23000\unit{s$^{-1}$}, compared with typically 32000\unit{s$^{-1}$} for \CaII{40} (a
reduction attributable to less efficient repumping from the D states).

\begin{figure}
\begin{centre}
\includegraphics[width=\onecolfig]{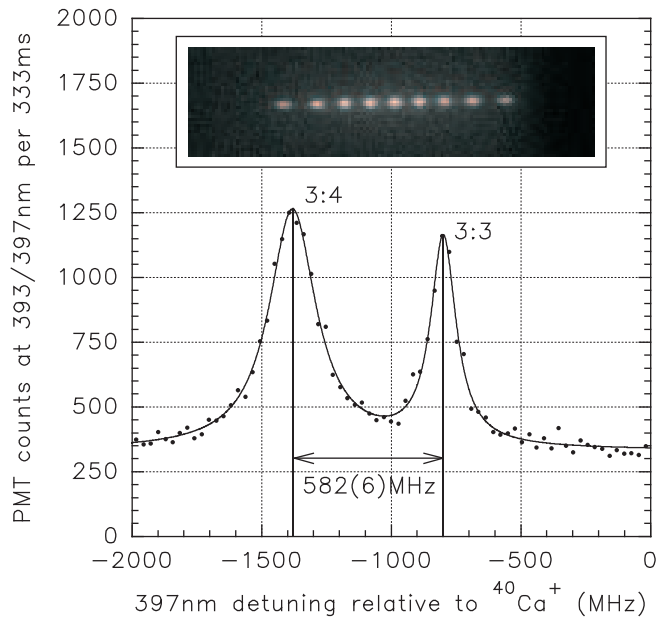}
\caption{%
Fluorescence signal from a single \CaII{43} ion as the frequency of the 397\nm\ ultraviolet repumping laser is scanned.
The 397\nm\ intensity was reduced to \ish 0.2\unit{mW/mm$^2$} for this scan, to reduce the widths of the peaks. A
double Lorentzian is fitted to the data: the separation between the two components is 582(6)\MHz\ (where a 1\% error is
allowed in deducing the frequency scale from a confocal cavity of 300\MHz\ free spectral range), in excellent agreement
with the known hyperfine splitting in the \lev{}{P}{1/2} level of 581\MHz. The
$\hfslev{S}{1/2}{3}\leftrightarrow\hfslev{P}{1/2}{4}$ component has been assigned the known shift of $-1381\MHz$
relative to \CaII{40}.
{\em Inset:\/} A pure crystal of nine \CaII{43} ions, loaded using periodic blocking of the 397\nm\ repumper to remove
even-isotope impurity ions, as described in the text. The separation between the closest ions is 13(1)\um.
}%
\label{F:ca43}
\end{centre}
\end{figure}

To confirm the identity of the first single ion we loaded by this method, we scanned the frequency of the 397\nm\
repumper: at low intensity, the two hyperfine components $\hfslev{S}{1/2}{3}\leftrightarrow\hfslev{P}{1/2}{4}$ and
$\hfslev{S}{1/2}{3}\leftrightarrow\hfslev{P}{1/2}{3}$ are well-resolved, figure~\ref{F:ca43}, and have the expected
separation. As a further check, the axial resonance frequency was measured and compared with that of \CaII{40}, giving
a mass number of 42.8(2).

If the photo-ionizing lasers and oven are left on for longer we find that we load crystals containing about 50\%
\CaII{43} and 50\% \CaII{40}. This may occur either because of charge-exchange, or because once \CaII{43} is trapped
and cooled, it can cool \CaII{40} sympathetically in spite of the heating effect of the 393\nm\ laser. We discovered
that it was possible to purify such mixed crystals by temporarily blocking the 397\nm\ ultraviolet repumper; when this
laser is blocked a small fluorescence signal is still visible from \CaII{43} (because of off-resonant repumping on
$\hfslev{S}{1/2}{3}\leftrightarrow\hfslev{P}{3/2}{2;3;4}$ by the 3\GHz-detuned 393\nm\ laser) but there is no longer
any laser capable of cooling \CaII{40} directly. We assist the \CaII{40} heating by switching on the 866\nm\ repumper,
tuned to be resonant with the \CaII{40} transition. One might expect the Coulomb coupling between the different species
to make this purification process rather inefficient, but we found on the contrary that it always seems successful at
removing the \CaII{40} ions without loss of any \CaII{43}. For example, a mixed crystal consisting of three \CaII{43}
ions and eight \CaII{40} ions was purified completely by blocking the 397\nm\ repumper for \ish 8\unit{s}. The
technique can also be applied during loading: with the oven and photo-ionizing lasers on, we watch the \CaII{43} ions
being loaded on the camera and, as soon as there is evidence of ``dark ions'' in the trap, block the 397\nm\ repumper
for a few seconds to purify the crystal, then un-block it to continue loading. In this way, pure \CaII{43} crystals of
arbitrary size can be loaded; the inset of figure~\ref{F:ca43} shows a 9-ion linear crystal.

\section{Photo-ionization and electron bombardment compared}

\begin{figure}
\begin{centre}
\includegraphics[width=\onecolfig]{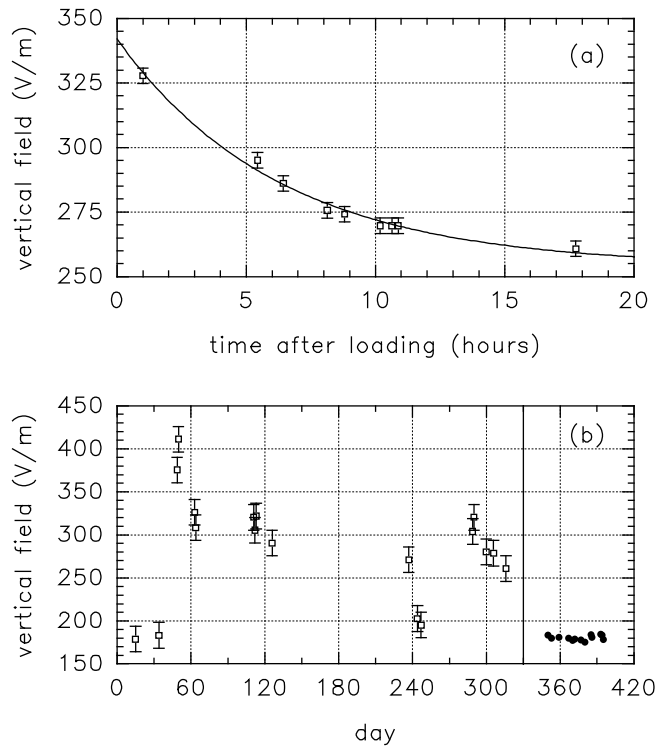}
\caption{%
(a)~Drift of the stray field versus time after loading with electron bombardment ionization. The magnitude of the
vertical field required to compensate horizontal ion micromotion is plotted. The micromotion compensation here was
performed by minimizing the width of the $\lev{}{S}{1/2}\leftrightarrow\lev{}{P}{1/2}$ spectrum; we assign an error of
$\pm 3\unit{V/m}$. An exponential decay function is fitted to the data (solid line); this has time-constant
6.3(7)\unit{hours} and amplitude 88(4)\unit{V/m}. No such drift is visible after loading by photo-ionization, at the
1\unit{V/m} level.
(b)~Variation of the steady-state stray field versus time, over the course of nearly a year for electron bombardment
loading (squares, left) and over the course of 45 days for photo-ionization loading (circles, right). Each point is
taken from a different load. For the electron bombardment data the error bar represents the estimated additional drift
possible after the final compensation measurement. For the photo-ionization data, the estimated error in compensating
the stray field is 1\unit{V/m}; the standard deviation of the data set is 2.8\unit{V/m}, indicating some residual
systematic variation (which may be due to the remaining atomic beam flux, or mechanical variations in the trap geometry
over time).
}%
\label{F:compensation}
\end{centre}
\end{figure}

The limitations of electron bombardment ionization are discussed in~\cite{01:Gulde}. A significant problem in our
apparatus was the effect of firing the electron gun on the stray electric fields in the trap. After loading, the static
field necessary to compensate the micromotion of the ions was found to decay exponentially with a time constant \ish
6\unit{hours}. For the typical load shown in figure~\ref{F:compensation}(a) the vertical stray field would not have
been stable to the 1\unit{V/m} resolution of our compensation method until \ish 28\unit{hours} after the trap was
loaded. We ascribe this drift to charge on insulating parts of the trap structure which decays with time, and have not
observed this systematic decay after photo-ionization loading. On longer timescales, between loads, we observed that
the vertical stray field changed by up to \ish 200\unit{V/m}; this is compared with recent photo-ionization
compensation data in figure~\ref{F:compensation}(b). The cause of these long-term changes in the stray field could have
been changes in the distribution of material deposited on the trap electrodes when the oven or electron gun was fired:
calcium deposited on the stainless steel r.f.\ electrodes could give rise to static fields of this magnitude, due to
the difference in work functions between these two metals.

This highlights a further issue: based on the work of the NIST Ion Storage group, we expect that material deposited
onto the trap electrodes from the oven and/or electron gun will contribute significantly to motional heating of trapped
ions in the sub-Doppler cooling regime relevant to quantum logic experiments~\cite{00:Turchette,02:Rowe}. The estimates
of calcium number density and temperature in the atomic beam given in section~\ref{S:oven423}, above, imply a flux of
$\ish 4\times 10^9\unit{atom/s/mm$^2$}$ on a surface normal to the atomic beam, at the 6\unit{A} oven current which was
necessary for loading the trap using electron bombardment. For the same loading rate, this can be reduced by over four
orders of magnitude using photo-ionization. Furthermore, the same photo-ionization loading rate can be maintained even
when using a well-collimated atomic beam designed to prevent deposition of atoms on the trap structure.

With maximum power in both photo-ionizing lasers, we measured a loading rate of 1.6(2)\unit{ion/s} at an oven current
of 3.25\unit{A}. This compares with an electron bombardment loading rate of \ish 1\unit{ion/s} at an oven current of
6.0\unit{A}. Extrapolating the number density data of figure~\ref{F:oven423}(a) to 6.0\unit{A} oven current indicates
that photo-ionization is a factor of $\ish 10^{4}$ times more efficient than our electron bombardment ionization setup.
A further two orders of magnitude enhancement of the loading efficiency in this apparatus would be available by (i)
increasing the spot sizes of the photo-ionizing lasers to reduce saturation effects on the 423\nm\ transition and (ii)
eliminating the optical fibre in the 389\nm\ beam. We note, however, that the per atom efficiency of photo-ionization
trap loading can be approached by techniques of surface ionization~\cite{00:Savard}.

The photo-ionization method overcomes the limitations of electron bombardment ionization: the species- and
isotope-selectivity ensure that unwanted ions are not trapped and furthermore allow studies of rare isotopes without
enriched sources; there is negligible effect due to the loading on the stray fields (at the level of \ish 1\unit{V/m}
in this trap); it works well irrespective of radial trap strength~\cite{01:Gulde}, so that compensation of stray fields
for different trap conditions is unnecessary; the high efficiency of photo-ionization means that the flux of the atomic
beam can be attenuated by many orders of magnitude which reduces immensely problems associated with deposition of
material on the trap structure. These advantages must be weighed against the complexities of two additional lasers,
compared with an electron gun inside the vacuum system.

\section{Conclusion}

We have presented quantitative studies of photo-ionization ion-trap loading for calcium ions, using the fluorescence
from the neutral atoms to estimate the absolute efficiency of this process. We find that, in our trap, this loading
method is around four orders of magnitude more efficient than conventional electron impact ionization, comparably to
Gulde \etal~\cite{01:Gulde}, and does not give rise to stray electric fields. We have shown that the setup of Gulde
\etal\ may be further simplified by eliminating frequency-stabilization for the 389\nm\ laser and, if high loading
rates are not required, by replacing this laser with a light-emitting diode. We have optimized the photo-ionization
procedure for the purposes of isotope-selection, demonstrating trapping of all the naturally occurring isotopes from a
natural abundance source. The rare species \CaII{43} and \CaII{46} have been laser-cooled for the first time.

The odd isotope \CaII{43} is of particular interest for experiments in quantum information processing and optical
frequency standards, and we have demonstrated the ability to load pure crystals of this ion. We note, however, that an
enriched source would still be advantageous for work with this isotope, since it would reduce any remaining deposition
of material from the atomic beam on the trap electrodes to the minimum possible. An interesting feature of
photo-ionization compared with electron bombardment ionization is that the laser beams allow it to be
spatially-selective: this could be useful for loading specific traps in proposed quantum computing architectures
involving microtrap arrays~\cite{00:Cirac}. The spatial selectivity could be extended to three dimensions by crossing
the 423\nm\ and 389\nm\ beams. The isotope-selectivity would also be invaluable for loading ``refrigerator''
ions~\cite{02:Kielpinski} in schemes where continuous cooling is provided by one isotope, while quantum logic
operations are carried out using another isotope: following the estimate in~\cite{02:Blinov}, the decoherence rate, due
to off-resonant photon scattering, of a \CaII{43} qubit being cooled sympathetically by a \CaII{42} refrigerator ion is
\ish 0.1/s for a trap with a heating rate of $10^3$ motional quanta per second~\cite{00:Turchette}.

Finally, in comparison with the work of Mortensen \etal, it is clear that the 272\nm\
4s$^2\,\lev{1}{S}{0}\leftrightarrow$ 4s5p\,\lev{1}{P}{1} transition is much more favourable for isotope-selective
ionization of calcium than the 423\nm\ transition used in this work, since it possesses a narrower natural width,
8.5(5)\MHz~\cite{93:Mitroy}, and isotope shifts about a factor of two larger~\cite{02:Mortensen}. However, this scheme
requires a significantly more complex laser source than the blue diode laser used here, and is somewhat less efficient
since it involves excitation deeper into the continuum. An unfortunate circumstance at the time of writing is the
unavailability of further diodes at 423\nm\ from Nichia Corporation but we are optimistic that, with blue laser diodes
under development by several other manufacturers and research laboratories, the situation will improve.

\begin{acknowledgments}
We are extremely grateful to Dr.\ Hugo van der Hart and Ms.\ Claire McKenna for performing calculations of the
photo-ionization cross-section. We would also like to thank the ion-trapping group of Prof.\ Dr.\ Rainer Blatt at
Universit\"{a}t Innsbruck for many useful discussions, Dr.\ Charles Donald for compiling the electron bombardment
ionization data in figure~\ref{F:compensation}, and Wolfgang Kemp at Toptica Photonics for helpful advice. Graham
Quelch contributed invaluable technical assistance. The research is supported by the EPSRC, ARDA
(P-43513-PH-QCO-02107-1) and the E.U. QGATES network. MM acknowledges the support of the Commonwealth Scholarship and
Fellowship Plan. DML is a Royal Society University Research Fellow and wishes to thank Dr.\ Bruce Warrington for {\tt
SevenSegmentDisplay} software development.
\end{acknowledgments}

\bibliography{calcium}

\begin{thebibliography}{35}
\expandafter\ifx\csname natexlab\endcsname\relax\def\natexlab#1{#1}\fi
\expandafter\ifx\csname bibnamefont\endcsname\relax
  \def\bibnamefont#1{#1}\fi
\expandafter\ifx\csname bibfnamefont\endcsname\relax
  \def\bibfnamefont#1{#1}\fi
\expandafter\ifx\csname citenamefont\endcsname\relax
  \def\citenamefont#1{#1}\fi
\expandafter\ifx\csname url\endcsname\relax
  \def\url#1{\texttt{#1}}\fi
\expandafter\ifx\csname urlprefix\endcsname\relax\def\urlprefix{URL }\fi
\providecommand{\bibinfo}[2]{#2}
\providecommand{\eprint}[2][]{\url{#2}}

\bibitem[{\citenamefont{Kj{\ae}rgaard et~al.}(2000)\citenamefont{Kj{\ae}rgaard,
  Hornek{\ae}r, Thommesen, Videsen, and Drewsen}}]{00:Kjaergaard}
\bibinfo{author}{\bibfnamefont{N.}~\bibnamefont{Kj{\ae}rgaard}},
  \bibinfo{author}{\bibfnamefont{L.}~\bibnamefont{Hornek{\ae}r}},
  \bibinfo{author}{\bibfnamefont{A.~M.} \bibnamefont{Thommesen}},
  \bibinfo{author}{\bibfnamefont{Z.}~\bibnamefont{Videsen}}, \bibnamefont{and}
  \bibinfo{author}{\bibfnamefont{M.}~\bibnamefont{Drewsen}},
  \bibinfo{journal}{Appl. Phys.} \textbf{\bibinfo{volume}{B71}},
  \bibinfo{pages}{207} (\bibinfo{year}{2000}).

\bibitem[{\citenamefont{Mortensen et~al.}(2002)\citenamefont{Mortensen,
  Lindballe, Jensen, Voigt, and Drewsen}}]{02:Mortensen}
\bibinfo{author}{\bibfnamefont{A.}~\bibnamefont{Mortensen}},
  \bibinfo{author}{\bibfnamefont{J.}~\bibnamefont{Lindballe}},
  \bibinfo{author}{\bibfnamefont{I.~S.} \bibnamefont{Jensen}},
  \bibinfo{author}{\bibfnamefont{D.}~\bibnamefont{Voigt}}, \bibnamefont{and}
  \bibinfo{author}{\bibfnamefont{M.}~\bibnamefont{Drewsen}}, in
  \emph{\bibinfo{booktitle}{XVIII International Conference on Atomic Physics
  (Poster Presentation Abstracts)}} (\bibinfo{publisher}{World Scientific,
  Singapore}, \bibinfo{year}{2002}), p. \bibinfo{pages}{253}.

\bibitem[{\citenamefont{Gulde et~al.}(2001)\citenamefont{Gulde, Rotter, Barton,
  Schmidt-Kaler, Blatt, and Hogervorst}}]{01:Gulde}
\bibinfo{author}{\bibfnamefont{S.}~\bibnamefont{Gulde}},
  \bibinfo{author}{\bibfnamefont{D.}~\bibnamefont{Rotter}},
  \bibinfo{author}{\bibfnamefont{P.}~\bibnamefont{Barton}},
  \bibinfo{author}{\bibfnamefont{F.}~\bibnamefont{Schmidt-Kaler}},
  \bibinfo{author}{\bibfnamefont{R.}~\bibnamefont{Blatt}}, \bibnamefont{and}
  \bibinfo{author}{\bibfnamefont{W.}~\bibnamefont{Hogervorst}},
  \bibinfo{journal}{Appl. Phys.} \textbf{\bibinfo{volume}{B73}},
  \bibinfo{pages}{861} (\bibinfo{year}{2001}).

\bibitem[{\citenamefont{Turchette et~al.}(2000)\citenamefont{Turchette,
  Kielpinski, King, Leibfried, Meekhof, Myatt, Rowe, Sackett, Wood, Itano
  et~al.}}]{00:Turchette}
\bibinfo{author}{\bibfnamefont{Q.~A.} \bibnamefont{Turchette}},
  \bibinfo{author}{\bibfnamefont{D.}~\bibnamefont{Kielpinski}},
  \bibinfo{author}{\bibfnamefont{B.~E.} \bibnamefont{King}},
  \bibinfo{author}{\bibfnamefont{D.}~\bibnamefont{Leibfried}},
  \bibinfo{author}{\bibfnamefont{D.~M.} \bibnamefont{Meekhof}},
  \bibinfo{author}{\bibfnamefont{C.~J.} \bibnamefont{Myatt}},
  \bibinfo{author}{\bibfnamefont{M.~A.} \bibnamefont{Rowe}},
  \bibinfo{author}{\bibfnamefont{C.~A.} \bibnamefont{Sackett}},
  \bibinfo{author}{\bibfnamefont{C.~S.} \bibnamefont{Wood}},
  \bibinfo{author}{\bibfnamefont{W.~M.} \bibnamefont{Itano}},
  \bibnamefont{et~al.}, \bibinfo{journal}{Phys. Rev.}
  \textbf{\bibinfo{volume}{A61}}, \bibinfo{pages}{063418}
  (\bibinfo{year}{2000}).

\bibitem[{\citenamefont{Mundt et~al.}(2003)\citenamefont{Mundt, Kreuter, Russo,
  Becher, Leibfried, Eschner, Schmidt-Kaler, and Blatt}}]{03:Mundt}
\bibinfo{author}{\bibfnamefont{A.~B.} \bibnamefont{Mundt}},
  \bibinfo{author}{\bibfnamefont{A.}~\bibnamefont{Kreuter}},
  \bibinfo{author}{\bibfnamefont{C.}~\bibnamefont{Russo}},
  \bibinfo{author}{\bibfnamefont{C.}~\bibnamefont{Becher}},
  \bibinfo{author}{\bibfnamefont{D.}~\bibnamefont{Leibfried}},
  \bibinfo{author}{\bibfnamefont{J.}~\bibnamefont{Eschner}},
  \bibinfo{author}{\bibfnamefont{F.}~\bibnamefont{Schmidt-Kaler}},
  \bibnamefont{and} \bibinfo{author}{\bibfnamefont{R.}~\bibnamefont{Blatt}},
  \bibinfo{journal}{Appl. Phys.} \textbf{\bibinfo{volume}{B76}},
  \bibinfo{pages}{117} (\bibinfo{year}{2003}).

\bibitem[{\citenamefont{Keller et~al.}(2003)\citenamefont{Keller, Lange,
  Hayasaka, Lange, and Walther}}]{03:Keller}
\bibinfo{author}{\bibfnamefont{M.}~\bibnamefont{Keller}},
  \bibinfo{author}{\bibfnamefont{B.}~\bibnamefont{Lange}},
  \bibinfo{author}{\bibfnamefont{K.}~\bibnamefont{Hayasaka}},
  \bibinfo{author}{\bibfnamefont{W.}~\bibnamefont{Lange}}, \bibnamefont{and}
  \bibinfo{author}{\bibfnamefont{H.}~\bibnamefont{Walther}},
  \bibinfo{journal}{Appl. Phys.} \textbf{\bibinfo{volume}{B76}},
  \bibinfo{pages}{125} (\bibinfo{year}{2003}).

\bibitem[{\citenamefont{Steane}(1997)}]{97:Steane}
\bibinfo{author}{\bibfnamefont{A.~M.} \bibnamefont{Steane}},
  \bibinfo{journal}{Appl. Phys.} \textbf{\bibinfo{volume}{B64}},
  \bibinfo{pages}{623} (\bibinfo{year}{1997}).

\bibitem[{\citenamefont{Lucas et~al.}(2003)\citenamefont{Lucas, Donald, Home,
  McDonnell, Ramos, Stacey, Stacey, Steane, and Webster}}]{03:Lucas}
\bibinfo{author}{\bibfnamefont{D.~M.} \bibnamefont{Lucas}},
  \bibinfo{author}{\bibfnamefont{C.~J.~S.} \bibnamefont{Donald}},
  \bibinfo{author}{\bibfnamefont{J.~P.} \bibnamefont{Home}},
  \bibinfo{author}{\bibfnamefont{M.~J.} \bibnamefont{McDonnell}},
  \bibinfo{author}{\bibfnamefont{A.}~\bibnamefont{Ramos}},
  \bibinfo{author}{\bibfnamefont{D.~N.} \bibnamefont{Stacey}},
  \bibinfo{author}{\bibfnamefont{J.-P.} \bibnamefont{Stacey}},
  \bibinfo{author}{\bibfnamefont{A.~M.} \bibnamefont{Steane}},
  \bibnamefont{and} \bibinfo{author}{\bibfnamefont{S.~C.}
  \bibnamefont{Webster}}, \bibinfo{journal}{Phil. Trans.}
  \textbf{\bibinfo{volume}{361}}, \bibinfo{pages}{1401} (\bibinfo{year}{2003}).

\bibitem[{\citenamefont{Schmidt-Kaler et~al.}(2003)\citenamefont{Schmidt-Kaler,
  Gulde, Riebe, Deuschle, Kreuter, Lancaster, Becher, Eschner, H{\"{a}}ffner,
  and Blatt}}]{03:SchmidtKaler}
\bibinfo{author}{\bibfnamefont{F.}~\bibnamefont{Schmidt-Kaler}},
  \bibinfo{author}{\bibfnamefont{S.}~\bibnamefont{Gulde}},
  \bibinfo{author}{\bibfnamefont{M.}~\bibnamefont{Riebe}},
  \bibinfo{author}{\bibfnamefont{T.}~\bibnamefont{Deuschle}},
  \bibinfo{author}{\bibfnamefont{A.}~\bibnamefont{Kreuter}},
  \bibinfo{author}{\bibfnamefont{G.}~\bibnamefont{Lancaster}},
  \bibinfo{author}{\bibfnamefont{C.}~\bibnamefont{Becher}},
  \bibinfo{author}{\bibfnamefont{J.}~\bibnamefont{Eschner}},
  \bibinfo{author}{\bibfnamefont{H.}~\bibnamefont{H{\"{a}}ffner}},
  \bibnamefont{and} \bibinfo{author}{\bibfnamefont{R.}~\bibnamefont{Blatt}},
  \bibinfo{journal}{J. Phys.} \textbf{\bibinfo{volume}{B36}},
  \bibinfo{pages}{623} (\bibinfo{year}{2003}).

\bibitem[{\citenamefont{Plumelle and Desaintfusien}(1993)}]{93:Plumelle}
\bibinfo{author}{\bibfnamefont{F.}~\bibnamefont{Plumelle}} \bibnamefont{and}
  \bibinfo{author}{\bibfnamefont{M.}~\bibnamefont{Desaintfusien}},
  \bibinfo{journal}{IEEE Trans. Intrum. Meas.} \textbf{\bibinfo{volume}{42}},
  \bibinfo{pages}{462} (\bibinfo{year}{1993}).

\bibitem[{\citenamefont{Boshier et~al.}(2000)\citenamefont{Boshier, Barwood,
  Huang, and Klein}}]{00:Boshier}
\bibinfo{author}{\bibfnamefont{M.~G.} \bibnamefont{Boshier}},
  \bibinfo{author}{\bibfnamefont{G.~P.} \bibnamefont{Barwood}},
  \bibinfo{author}{\bibfnamefont{G.}~\bibnamefont{Huang}}, \bibnamefont{and}
  \bibinfo{author}{\bibfnamefont{H.~A.} \bibnamefont{Klein}},
  \bibinfo{journal}{Appl. Phys.} \textbf{\bibinfo{volume}{B71}},
  \bibinfo{pages}{51} (\bibinfo{year}{2000}).

\bibitem[{\citenamefont{Cline}()}]{PC:OakRidge}
\bibinfo{author}{\bibfnamefont{R.~L.} \bibnamefont{Cline}},
  \bibinfo{note}{{(Isotope Distribution, Oak Ridge National Laboratory) Private
  communication}}.

\bibitem[{\citenamefont{Alheit et~al.}(1996)\citenamefont{Alheit, Enders, and
  Werth}}]{96:Alheit}
\bibinfo{author}{\bibfnamefont{R.}~\bibnamefont{Alheit}},
  \bibinfo{author}{\bibfnamefont{K.}~\bibnamefont{Enders}}, \bibnamefont{and}
  \bibinfo{author}{\bibfnamefont{G.}~\bibnamefont{Werth}},
  \bibinfo{journal}{Appl. Phys.} \textbf{\bibinfo{volume}{B62}},
  \bibinfo{pages}{511} (\bibinfo{year}{1996}).

\bibitem[{\citenamefont{Toyoda et~al.}(2001)\citenamefont{Toyoda, Kataoka, Kai,
  Miura, Watanabe, and Urabe}}]{01:Toyoda}
\bibinfo{author}{\bibfnamefont{K.}~\bibnamefont{Toyoda}},
  \bibinfo{author}{\bibfnamefont{H.}~\bibnamefont{Kataoka}},
  \bibinfo{author}{\bibfnamefont{Y.}~\bibnamefont{Kai}},
  \bibinfo{author}{\bibfnamefont{A.}~\bibnamefont{Miura}},
  \bibinfo{author}{\bibfnamefont{M.}~\bibnamefont{Watanabe}}, \bibnamefont{and}
  \bibinfo{author}{\bibfnamefont{S.}~\bibnamefont{Urabe}},
  \bibinfo{journal}{Appl. Phys.} \textbf{\bibinfo{volume}{B72}},
  \bibinfo{pages}{327} (\bibinfo{year}{2001}).

\bibitem[{\citenamefont{N{\"{o}}rtersh{\"{a}}user
  et~al.}(1998{\natexlab{a}})\citenamefont{N{\"{o}}rtersh{\"{a}}user, Blaum,
  Icker, M{\"{u}}ller, Schmitt, Wendt, and Wiche}}]{98a:Nortershauser}
\bibinfo{author}{\bibfnamefont{W.}~\bibnamefont{N{\"{o}}rtersh{\"{a}}user}},
  \bibinfo{author}{\bibfnamefont{K.}~\bibnamefont{Blaum}},
  \bibinfo{author}{\bibfnamefont{K.}~\bibnamefont{Icker}},
  \bibinfo{author}{\bibfnamefont{P.}~\bibnamefont{M{\"{u}}ller}},
  \bibinfo{author}{\bibfnamefont{A.}~\bibnamefont{Schmitt}},
  \bibinfo{author}{\bibfnamefont{K.}~\bibnamefont{Wendt}}, \bibnamefont{and}
  \bibinfo{author}{\bibfnamefont{B.}~\bibnamefont{Wiche}},
  \bibinfo{journal}{Eur. Phys. J.} \textbf{\bibinfo{volume}{D2}},
  \bibinfo{pages}{33} (\bibinfo{year}{1998}{\natexlab{a}}).

\bibitem[{\citenamefont{N{\"{o}}rtersh{\"{a}}user
  et~al.}(1998{\natexlab{b}})\citenamefont{N{\"{o}}rtersh{\"{a}}user,
  Trautmann, Wendt, and Bushaw}}]{98b:Nortershauser}
\bibinfo{author}{\bibfnamefont{W.}~\bibnamefont{N{\"{o}}rtersh{\"{a}}user}},
  \bibinfo{author}{\bibfnamefont{N.}~\bibnamefont{Trautmann}},
  \bibinfo{author}{\bibfnamefont{K.}~\bibnamefont{Wendt}}, \bibnamefont{and}
  \bibinfo{author}{\bibfnamefont{B.~A.} \bibnamefont{Bushaw}},
  \bibinfo{journal}{Spectrochimica Acta} \textbf{\bibinfo{volume}{B53}},
  \bibinfo{pages}{709} (\bibinfo{year}{1998}{\natexlab{b}}).

\bibitem[{\citenamefont{Mitroy}(1993)}]{93:Mitroy}
\bibinfo{author}{\bibfnamefont{J.}~\bibnamefont{Mitroy}}, \bibinfo{journal}{J.
  Phys.} \textbf{\bibinfo{volume}{B26}}, \bibinfo{pages}{3703}
  (\bibinfo{year}{1993}).

\bibitem[{\citenamefont{Barton et~al.}(2000)\citenamefont{Barton, Donald,
  Lucas, Stevens, Steane, and Stacey}}]{00:Barton}
\bibinfo{author}{\bibfnamefont{P.}~\bibnamefont{Barton}},
  \bibinfo{author}{\bibfnamefont{C.~J.~S.} \bibnamefont{Donald}},
  \bibinfo{author}{\bibfnamefont{D.~M.} \bibnamefont{Lucas}},
  \bibinfo{author}{\bibfnamefont{D.~A.} \bibnamefont{Stevens}},
  \bibinfo{author}{\bibfnamefont{A.~M.} \bibnamefont{Steane}},
  \bibnamefont{and} \bibinfo{author}{\bibfnamefont{D.~N.}
  \bibnamefont{Stacey}}, \bibinfo{journal}{Phys. Rev.}
  \textbf{\bibinfo{volume}{A62}}, \bibinfo{pages}{032503}
  (\bibinfo{year}{2000}).

\bibitem[{\citenamefont{Coplen et~al.}(2002)\citenamefont{Coplen, B{\"{o}}hlke,
  Bi{\`e}vre, Ding, Holden, Hopple, Krouse, Lamberty, Peiser, R{\'e}v{\'e}sz
  et~al.}}]{02:Coplen}
\bibinfo{author}{\bibfnamefont{T.~B.} \bibnamefont{Coplen}},
  \bibinfo{author}{\bibfnamefont{J.~K.} \bibnamefont{B{\"{o}}hlke}},
  \bibinfo{author}{\bibfnamefont{P.~D.} \bibnamefont{Bi{\`e}vre}},
  \bibinfo{author}{\bibfnamefont{T.}~\bibnamefont{Ding}},
  \bibinfo{author}{\bibfnamefont{N.~E.} \bibnamefont{Holden}},
  \bibinfo{author}{\bibfnamefont{J.~A.} \bibnamefont{Hopple}},
  \bibinfo{author}{\bibfnamefont{H.~R.} \bibnamefont{Krouse}},
  \bibinfo{author}{\bibfnamefont{A.}~\bibnamefont{Lamberty}},
  \bibinfo{author}{\bibfnamefont{H.~S.} \bibnamefont{Peiser}},
  \bibinfo{author}{\bibfnamefont{K.}~\bibnamefont{R{\'e}v{\'e}sz}},
  \bibnamefont{et~al.}, \bibinfo{journal}{Pure Appl. Chem.}
  \textbf{\bibinfo{volume}{74}}, \bibinfo{pages}{1987} (\bibinfo{year}{2002}).

\bibitem[{\citenamefont{M{\aa}rtensson-Pendrill
  et~al.}(1992)\citenamefont{M{\aa}rtensson-Pendrill, Ynnerman, Warston,
  Vermeeren, Silverans, Klein, Neugart, Schulz, Lievens, and the
  ISOLDE~Collaboration}}]{92:Martensson}
\bibinfo{author}{\bibfnamefont{A.~M.} \bibnamefont{M{\aa}rtensson-Pendrill}},
  \bibinfo{author}{\bibfnamefont{A.}~\bibnamefont{Ynnerman}},
  \bibinfo{author}{\bibfnamefont{H.}~\bibnamefont{Warston}},
  \bibinfo{author}{\bibfnamefont{L.}~\bibnamefont{Vermeeren}},
  \bibinfo{author}{\bibfnamefont{R.~E.} \bibnamefont{Silverans}},
  \bibinfo{author}{\bibfnamefont{A.}~\bibnamefont{Klein}},
  \bibinfo{author}{\bibfnamefont{R.}~\bibnamefont{Neugart}},
  \bibinfo{author}{\bibfnamefont{C.}~\bibnamefont{Schulz}},
  \bibinfo{author}{\bibfnamefont{P.}~\bibnamefont{Lievens}}, \bibnamefont{and}
  \bibinfo{author}{\bibnamefont{the ISOLDE~Collaboration}},
  \bibinfo{journal}{Phys. Rev.} \textbf{\bibinfo{volume}{A45}},
  \bibinfo{pages}{4675} (\bibinfo{year}{1992}).

\bibitem[{\citenamefont{Berkeland et~al.}(1998)\citenamefont{Berkeland, Miller,
  Bergquist, Itano, and Wineland}}]{98:Berkeland}
\bibinfo{author}{\bibfnamefont{D.~J.} \bibnamefont{Berkeland}},
  \bibinfo{author}{\bibfnamefont{J.~D.} \bibnamefont{Miller}},
  \bibinfo{author}{\bibfnamefont{J.~C.} \bibnamefont{Bergquist}},
  \bibinfo{author}{\bibfnamefont{W.~M.} \bibnamefont{Itano}}, \bibnamefont{and}
  \bibinfo{author}{\bibfnamefont{D.~J.} \bibnamefont{Wineland}},
  \bibinfo{journal}{J. Appl. Phys.} \textbf{\bibinfo{volume}{83}},
  \bibinfo{pages}{5025} (\bibinfo{year}{1998}).

\bibitem[{\citenamefont{Demtr{\"o}der}(1982)}]{Bk:Demtroder}
\bibinfo{author}{\bibfnamefont{W.}~\bibnamefont{Demtr{\"o}der}},
  \emph{\bibinfo{title}{Laser Spectroscopy}}
  (\bibinfo{publisher}{Springer-Verlag}, \bibinfo{year}{1982}).

\bibitem[{\citenamefont{Smith}(1972)}]{72:Smith}
\bibinfo{author}{\bibfnamefont{G.}~\bibnamefont{Smith}}, \bibinfo{journal}{J.
  Phys.} \textbf{\bibinfo{volume}{B5}}, \bibinfo{pages}{2310}
  (\bibinfo{year}{1972}).

\bibitem[{\citenamefont{Scoles}(1988)}]{Bk:Scoles}
\bibinfo{editor}{\bibfnamefont{G.}~\bibnamefont{Scoles}}, ed.,
  \emph{\bibinfo{title}{Atomic and Molecular Beam Methods}}
  (\bibinfo{publisher}{Oxford University Press}, \bibinfo{year}{1988}).

\bibitem[{\citenamefont{Barin}(1993)}]{Bk:Barin}
\bibinfo{author}{\bibfnamefont{I.}~\bibnamefont{Barin}},
  \emph{\bibinfo{title}{Thermochemical Data of Pure Substances}}
  (\bibinfo{publisher}{VCH, Weinheim}, \bibinfo{year}{1993}),
  \bibinfo{note}{(Second edition)}.

\bibitem[{\citenamefont{van~der Hart and McKenna}()}]{PC:vanderHart}
\bibinfo{author}{\bibfnamefont{H.~W.} \bibnamefont{van~der Hart}}
  \bibnamefont{and} \bibinfo{author}{\bibfnamefont{C.}~\bibnamefont{McKenna}},
  \bibinfo{note}{{Private communication}}.

\bibitem[{\citenamefont{Kuhn}(1969)}]{Bk:Kuhn}
\bibinfo{author}{\bibfnamefont{H.}~\bibnamefont{Kuhn}},
  \emph{\bibinfo{title}{Atomic Spectra}} (\bibinfo{publisher}{Longmans,
  London}, \bibinfo{year}{1969}).

\bibitem[{\citenamefont{Cowan}(1981)}]{Bk:Cowan}
\bibinfo{author}{\bibfnamefont{R.~D.} \bibnamefont{Cowan}},
  \emph{\bibinfo{title}{The Theory of Atomic Structure and Spectra}}
  (\bibinfo{publisher}{California Press}, \bibinfo{year}{1981}).

\bibitem[{\citenamefont{Blinov et~al.}(2002)\citenamefont{Blinov, Deslauriers,
  Lee, Madsen, Miller, and Monroe}}]{02:Blinov}
\bibinfo{author}{\bibfnamefont{B.~B.} \bibnamefont{Blinov}},
  \bibinfo{author}{\bibfnamefont{L.}~\bibnamefont{Deslauriers}},
  \bibinfo{author}{\bibfnamefont{P.}~\bibnamefont{Lee}},
  \bibinfo{author}{\bibfnamefont{M.~J.} \bibnamefont{Madsen}},
  \bibinfo{author}{\bibfnamefont{R.}~\bibnamefont{Miller}}, \bibnamefont{and}
  \bibinfo{author}{\bibfnamefont{C.}~\bibnamefont{Monroe}},
  \bibinfo{journal}{Phys. Rev.} \textbf{\bibinfo{volume}{A65}},
  \bibinfo{pages}{040304} (\bibinfo{year}{2002}).

\bibitem[{\citenamefont{Bowe et~al.}(1999)\citenamefont{Bowe, Hornek{\ae}r,
  Brodersen, Drewsen, Hangst, and Schiffer}}]{99:Bowe}
\bibinfo{author}{\bibfnamefont{P.}~\bibnamefont{Bowe}},
  \bibinfo{author}{\bibfnamefont{L.}~\bibnamefont{Hornek{\ae}r}},
  \bibinfo{author}{\bibfnamefont{C.}~\bibnamefont{Brodersen}},
  \bibinfo{author}{\bibfnamefont{M.}~\bibnamefont{Drewsen}},
  \bibinfo{author}{\bibfnamefont{J.~S.} \bibnamefont{Hangst}},
  \bibnamefont{and} \bibinfo{author}{\bibfnamefont{J.~P.}
  \bibnamefont{Schiffer}}, \bibinfo{journal}{Phys. Rev. Lett.}
  \textbf{\bibinfo{volume}{82}}, \bibinfo{pages}{2071} (\bibinfo{year}{1999}).

\bibitem[{\citenamefont{Siemers et~al.}(1992)\citenamefont{Siemers, Schubert,
  Blatt, Neuhauser, and Toschek}}]{92:Siemers}
\bibinfo{author}{\bibfnamefont{I.}~\bibnamefont{Siemers}},
  \bibinfo{author}{\bibfnamefont{M.}~\bibnamefont{Schubert}},
  \bibinfo{author}{\bibfnamefont{R.}~\bibnamefont{Blatt}},
  \bibinfo{author}{\bibfnamefont{W.}~\bibnamefont{Neuhauser}},
  \bibnamefont{and} \bibinfo{author}{\bibfnamefont{P.~E.}
  \bibnamefont{Toschek}}, \bibinfo{journal}{Eur. Phys. Lett.}
  \textbf{\bibinfo{volume}{18}}, \bibinfo{pages}{139} (\bibinfo{year}{1992}).

\bibitem[{\citenamefont{Rowe et~al.}(2002)\citenamefont{Rowe, Ben-Kish,
  DeMarco, Leibfried, Meyer, Beall, Britton, Hughes, Itano, Jelenkovi\'{c}
  et~al.}}]{02:Rowe}
\bibinfo{author}{\bibfnamefont{M.~A.} \bibnamefont{Rowe}},
  \bibinfo{author}{\bibfnamefont{A.}~\bibnamefont{Ben-Kish}},
  \bibinfo{author}{\bibfnamefont{B.}~\bibnamefont{DeMarco}},
  \bibinfo{author}{\bibfnamefont{D.}~\bibnamefont{Leibfried}},
  \bibinfo{author}{\bibfnamefont{V.}~\bibnamefont{Meyer}},
  \bibinfo{author}{\bibfnamefont{J.}~\bibnamefont{Beall}},
  \bibinfo{author}{\bibfnamefont{J.}~\bibnamefont{Britton}},
  \bibinfo{author}{\bibfnamefont{J.}~\bibnamefont{Hughes}},
  \bibinfo{author}{\bibfnamefont{W.~M.} \bibnamefont{Itano}},
  \bibinfo{author}{\bibfnamefont{B.}~\bibnamefont{Jelenkovi\'{c}}},
  \bibnamefont{et~al.}, \bibinfo{journal}{Quant. Inform. Comp.}
  \textbf{\bibinfo{volume}{2}}, \bibinfo{pages}{257} (\bibinfo{year}{2002}).

\bibitem[{\citenamefont{Savard and Werth}(2000)}]{00:Savard}
\bibinfo{author}{\bibfnamefont{G.}~\bibnamefont{Savard}} \bibnamefont{and}
  \bibinfo{author}{\bibfnamefont{G.}~\bibnamefont{Werth}},
  \bibinfo{journal}{Annu. Rev. Nucl. Part. Sci.} \textbf{\bibinfo{volume}{50}},
  \bibinfo{pages}{119} (\bibinfo{year}{2000}).

\bibitem[{\citenamefont{Cirac and Zoller}(2000)}]{00:Cirac}
\bibinfo{author}{\bibfnamefont{J.~I.} \bibnamefont{Cirac}} \bibnamefont{and}
  \bibinfo{author}{\bibfnamefont{P.}~\bibnamefont{Zoller}},
  \bibinfo{journal}{Nature} \textbf{\bibinfo{volume}{404}},
  \bibinfo{pages}{579} (\bibinfo{year}{2000}).

\bibitem[{\citenamefont{Kielpinski et~al.}(2002)\citenamefont{Kielpinski,
  Monroe, and Wineland}}]{02:Kielpinski}
\bibinfo{author}{\bibfnamefont{D.}~\bibnamefont{Kielpinski}},
  \bibinfo{author}{\bibfnamefont{C.}~\bibnamefont{Monroe}}, \bibnamefont{and}
  \bibinfo{author}{\bibfnamefont{D.~J.} \bibnamefont{Wineland}},
  \bibinfo{journal}{Nature} \textbf{\bibinfo{volume}{417}},
  \bibinfo{pages}{709} (\bibinfo{year}{2002}).

\end{thebibliography}
\end{document}